\documentclass[showpacs,twocolumn,aps,pra,longbibliography,superscriptaddress,notitlepage]{revtex4-1}
\pdfoutput=1
\usepackage{natbib}
\usepackage{graphicx} \usepackage{hyperref}
\usepackage{tikz}
\usetikzlibrary{shadings,decorations.markings,fadings}
\usepackage{pgfplots}
\usepackage{quantikz}
\usetikzlibrary{backgrounds,fit,decorations.pathreplacing,calc}
\usepackage{commath,amsmath,amssymb}
\usepackage{braket}
\usepackage[capitalize]{cleveref}
\usepackage{soul}
\newcommand\lseq[1]{\stackrel{{\rm LS}}{=}}

\newcommand{\Tr}{{\rm Tr}}

\DeclareRobustCommand{\rchi}{{\mathpalette\irchi\relax}}
\newcommand{\irchi}[2]{\raisebox{\depth}{$#1\chi$}}

\newcommand{\Dmn}{\Delta_{mn}}
\newcommand{\Dmnp}{\Delta_{m'n'}}
\newcommand{\Ga}{\Gamma}
\newcommand{\gaij}{\gamma_{ij}}

\begin{document}
\title{Simple Diagonal State Designs with Reconfigurable Real-Time Circuits}
\author{Yizhi Shen}
\email{yizhis@lbl.gov}
\affiliation{Applied Mathematics and Computational Research Division, Lawrence Berkeley National Laboratory, Berkeley, CA 94720, USA}

\author{Katherine Klymko}
\affiliation{NERSC, Lawrence Berkeley National Laboratory, Berkeley, CA 94720, USA}

\author{Eran Rabani}
\affiliation{Materials Sciences Division, Lawrence Berkeley National Laboratory, Berkeley, CA 94720, USA}
\affiliation{Department of Chemistry, University of California, Berkeley, Berkeley, CA 94720, USA}
\affiliation{The Sackler Center for Computational Molecular and Materials Science, Tel Aviv University, Tel Aviv 69978, Israel}

\author{Norm M. Tubman}
 \affiliation{NASA Ames Research Center, Moffett Field, CA 94035, USA}

\author{Daan Camps}
\affiliation{NERSC, Lawrence Berkeley National Laboratory, Berkeley, CA 94720, USA}

\author{Roel Van Beeumen}
\affiliation{Applied Mathematics and Computational Research Division, Lawrence Berkeley National Laboratory, Berkeley, CA 94720, USA}

\author{Michael Lindsey}
\affiliation{Department of Mathematics, University of California, Berkeley, Berkeley, CA 94720, USA}

\begin{abstract}
Unitary designs are widely used in quantum computation, but in many practical settings it suffices to construct a diagonal state design generated with unitary gates diagonal in the computational basis. In this work, we introduce a simple and efficient diagonal state 3-design based on real-time evolutions under 2-local Hamiltonians. Our construction is inspired by the classical Girard-Hutchinson trace estimator in that it involves the stochastic preparation of many random-phase states. Though the exact Girard-Hutchinson states are not tractably implementable on a quantum computer, we can construct states that match the statistical moments of the Girard-Hutchinson states with real-time evolution. Importantly, our random states are all generated using the same Hamiltonians for real-time evolution, with the randomness arising solely from stochastic variations in the durations of the evolutions. In this sense, the circuit is fully reconfigurable and thus suited for near-term realizations on both digital and analog platforms. Moreover, we show how to extend our construction to achieve diagonal state designs of arbitrarily high order.
\end{abstract}

\maketitle
%%%% Introduction %%%% 
\section{Introduction}

Unitary designs, which faithfully reproduce statistical moments of the Haar distribution, 
play an important role in quantum information science, with wide-ranging applications including channel benchmarking, state or process tomography, and quantum linear algebra~\cite{Hayden2004,Fidelity,Preskill_shadow,Mele2024introductiontohaar}. For many problems, it suffices to consider the class of diagonal unitary designs, which consist solely of diagonal unitary operations~\cite{diagonal_2013,diagonal_2014}. In principle, constructing a diagonal design should simplify the resulting circuits. Moreover, the restriction to purely diagonal gates augments the feasibility of fault-tolerant, error-corrected implementations~\cite{Aliferis_2009}. Despite their simplicity, diagonal designs can be highly useful in randomization tasks. Notably, their scrambling properties are sufficient for powerful tomographic probes that predict arbitrary state properties~\cite{Hu2022,Tran2023,McGinley2023,liu2024predicting}.

In this manuscript, we introduce a quantum algorithm for constructing a diagonal 3-design, which samples random states within a quantum many-body Hilbert space. To prepare these random states (i.e., wavefunctions), we make use of real-time evolutions under a fixed set of simple and commuting auxiliary Hamiltonians, where only the evolution durations are randomized (and furthermore independent of the system size). By phase cancellation, the random states yield a desired stochastic resolution of identity. Notably, such randomized real-time evolutions, formally viewed as a sequence of commuting Pauli exponentials, enable a shallow implementation via gadget- or measurement-based rules~\cite{phasegadget_2019,Litinski2019gameofsurfacecodes,moflic2024constantdepthimplementationpauli}. For example, according to the phase gadget rule, each diagonal Pauli exponential can be decomposed into a single-qubit $Z$ rotation and two ladders of CNOT gates. Since our fixed auxiliary Hamiltonians are conveniently 2-local and only the evolution durations are randomized, we say that the circuit implementing our diagonal 3-design is \textit{reconfigurable}. That is, across a variety of both digital and analog platforms, we construct a single hardware-native circuit designed for reuse over many samples.

Compared to other approaches for constructing state designs, the strengths of our approach are that it (1) generates random states through simple real-time evolutions whose durations can be randomized independently of the system size, (2) exactly attains the desired statistical moments while avoiding expensive handling of exponentially large Hilbert spaces, and (3) significantly enhances practical near-term quantum efficiency, since it eliminates the overhead required for approaches that implement a distinct circuit for each sampling instance.

The manuscript is organized as follows. In \cref{sec:theory}, we start with an overview of unitary and diagonal unitary designs, highlighting the connections and disparities. We then expand on the intuition that real-time evolutions of randomized duration can fulfill the conditions of diagonal design. In \cref{sec:main results}, we present a practical approach for constructing a diagonal design. We explicitly compute its moments and detail its reconfigurable circuit representations on both digital and analog platforms. In \cref{sec:conclusion}, we summarize and conclude.

The real-time construction introduced in \cref{sec:main results} is well-suited for state tomography, in particular for classical shadow protocols. Interestingly, recent ideas based on Hamiltonian-driven shadows~\cite{Hu2022,Tran2023,McGinley2023,liu2024predicting} have demonstrated potential and permit native implementations on analog simulators. In particular, real-time evolution following a single random duration (discussed in detail within \cref{subsec:aside}) can be employed to generate shadows as considered in~\cite{liu2024predicting}, which assumes access to the ideal random diagonal ensemble. Our work reveals that this ensemble can in fact be sampled efficiently using several independent time evolutions of tractable duration.

% The real-time construction in \cref{sec:main results} is well-suited for central randomization tasks in quantum computation. Notably, the state scrambling property of our diagonal 3-design can be directly translated into tomographic probes helpful for predicting observables. Recent ideas based on Hamiltonian-driven shadows~\cite{Hu2022,Tran2023,McGinley2023,liu2024predicting} have demonstrated promising potential due to their native implementations on analog simulators. These approaches leverage the inherent dynamics of quantum systems, offering an efficient means to generate the desired randomness. As an example, real-time evolution following a single random duration (discussed concretely in \cref{subsec:aside}) can be used to generate classical shadows considered in the work~\cite{liu2024predicting}, which assumes access to the ideal random diagonal ensemble. 

\begin{figure*}[tbh!]
    \includegraphics[scale=0.485]{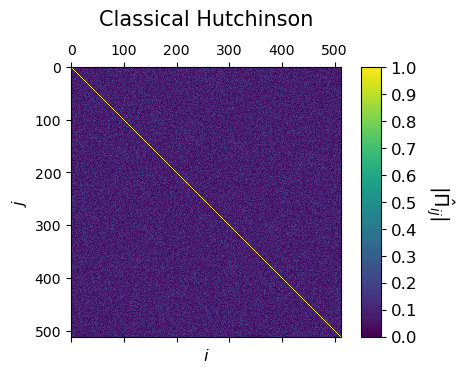}
    \includegraphics[scale=0.485]{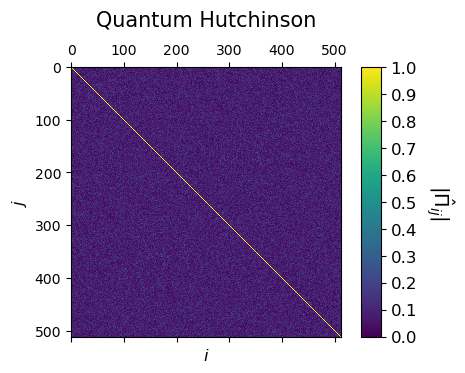}
    \includegraphics[scale=0.485]{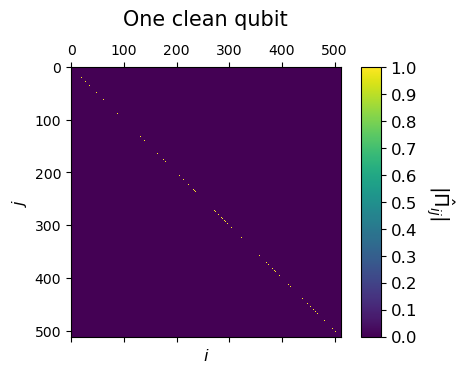}
    \caption{Approximate resolution of identity $\hat{\Pi} = \frac{N}{K} \sum_{k=1}^{K} \ket{\rchi_k} \bra{\rchi_k}$ constructed with $K=10^2$ random states of various types. \textbf{Left}: Classical Hutchinson states. \textbf{Center}: Quantum Hutchinson states (this work). \textbf{Right}: Uniformly sampled computational basis, corresponding to the one clean qubit construction. Matrix elements $\hat{\Pi}_{ij}$ of the approximate identity operator are colored by their magnitudes. Color interpolates linearly from dark blue ($\lvert \hat{\Pi}_{ij}\rvert = 0$) to yellow ($\lvert \hat{\Pi}_{ij}\rvert = 1$). }
    \label{fig:QH}
\end{figure*}

%%%% Theory %%%% 
\section{\label{sec:theory}Review of unitary and diagonal state designs}

A unitary design is an ensemble of unitary operators $U \sim p$ drawn according to a distribution $p$ which matches certain statistical moments of the Haar distribution over the unitary group. Specifically, a unitary $d$-design fulfills the defining property that for any quantum state $\ket{\psi}$,
\begin{align}
    \mathbb{E}_{U \sim p} \left[ \big( U \ket{\psi} \bra{\psi} U^{\dagger} \big)^{\otimes d} \right] = \mathbb{E}_{U \sim \rm Haar} \left[ \big(U \ket{\psi} \bra{\psi} U^{\dagger} \big)^{\otimes d} \right].
    \label{eq:d-design}
\end{align}
The matching property of \cref{eq:d-design} is valuable for a variety of tasks due to the prohibitive cost of explicitly sampling a random quantum circuit from the Haar distribution. In contrast, unitary designs offer a more feasible alternative as they are comparatively simpler to construct.~\cite{Fidelity,Gross2007,Harrow2009}

Meanwhile, \emph{diagonal} unitary designs relax the requirement of matching the full Haar randomness. As its name suggests, a diagonal design involves the action of diagonal unitary operators $U \sim p_{\rm diag}$ on some reference state~\cite{diagonal_2013,diagonal_2014}. To introduce the associated moment-matching property analogous to \cref{eq:d-design}, we first consider the $N$-dimensional random-phase state $\ket{\phi} \in \mathbb{C}^N$,
\begin{align}
    \ket{\phi} = \sum_{\Vec{n} \in \{0,1\}^{Q}  }^{2^{Q}~{\rm bitstrings} } \phi_{\Vec{n}} \ket{\Vec{n}} = \frac{1}{\sqrt{N}} \sum_{\Vec{n} \in \{0,1\}^{Q}  } e^{-i \theta_{\Vec{n} } } \ket{\Vec{n}},
    \label{eq:hutch}
\end{align}
where we allocate $Q = \lceil \log_2 N \rceil$ qubits to store the state, and $\theta_{\Vec{n} } \sim  \mathrm{Uniform}(0,2\pi)$ are independent and identically distributed random variables for all computational basis elements $\Vec{n}$. Historically, the use of random-phase vectors $\ket{\phi}$ can be traced back to the seminal work of Girard and Hutchinson on stochastic trace estimation~\cite{girard1989fast,Hutch,random_phase}, for which the coefficients $\phi_{\Vec{n}}$ are sampled independently and uniformly from $\{ \pm 1 \}$, i.e., the Rademacher distribution, on a classical computer. For this reason, throughout our manuscript we refer to the state $\vert \phi \rangle $ following \cref{eq:hutch} as a classical or conventional Hutchinson state. Accordingly, a diagonal $d$-design is an ensemble of diagonal unitaries such that for some reference state $\ket{\rchi_0}$, it holds that 
\begin{align}
    \mathbb{E}_{U \sim p_{\rm diag}} \left[ \big( U \ket{\rchi_0} \bra{\rchi_0} U^{\dagger} \big)^{\otimes d} \right] = \mathbb{E} \left[ \big(\ket{\phi} \bra{\phi}\big)^{\otimes d} \right],
    \label{eq:diag d-design}
\end{align}
where $\vert \phi \rangle$ on the right-hand side of \cref{eq:diag d-design} is understood to be a conventional Hutchinson state as in \cref{eq:hutch}. We comment that it is not tractable to sample wavefunctions with independent components in the computational basis, such as the conventional Hutchinson state, as the cost grows exponentially in the system size.

The matching property prescribed above can be leveraged in computational routines essential for various quantum algorithms. Among the earliest applications is the one clean qubit model of computation \cite{one_clean_qubit}, a computing paradigm targeting maximally mixed states. In particular, the state $\frac{1}{N} \mathbb{I}$ is stochastically simulated by uniformly sampling the computational basis $\{\ket{\Vec{n}}\}_{n=0}^{N-1}$, which forms a 1-design. Moreover, the application of diagonal $2$-designs to trace estimation matches the variance yielded by the conventional Hutchinson estimator, which is optimal among all estimators with independent components in the computational basis~\cite{random_phase}. The connection between diagonal designs and trace estimation has not been previously observed to our knowledge. In addition, diagonal 3-designs can be exploited to perform accurate state tomography. This capability is crucial for recent Hamiltonian-driven shadows~\cite{Hu2022,Tran2023,McGinley2023,liu2024predicting}, which have demonstrated efficient prediction of arbitrary state observables on practical analog simulators.

To construct a state design, typically the scrambling circuits are composed randomly from some suitable pool of candidate gates. For example, the Hadamard, phase, and CNOT gates combinatorially generate the Clifford group as a unitary 2-design, while single-qubit and controlled rotations \cite{diagonal_2013} are sufficient to generate a diagonal unitary 2-design. These approaches hence necessitate the realization of a distinct circuit for each sampled random state. Despite encouraging theoretical guarantees on circuit preparation such as polynomial gate count~\cite{diagonal_2013,diagonal_2014,Cleve2016}, implementing these designs on near-term hardware would still be costly, since the generation of each random state requires executing a new circuit compilation with varying gate composition and sequencing.

In our approach, we prioritize minimal quantum resource requirements for constructing a diagonal design. 
The central ingredient is the real-time evolution of a single, easily-prepared reference state for a random duration of time under an efficiently simulable Hamiltonian. Using this ingredient, we can generate random states achieving the matching identities~\cref{eq:diag d-design} up to $d = 3$, later extending to the case $d \geq 4$. Importantly, our diagonal design is reconfigurable in the sense that we have described above (as it bypasses the repetitive synthesis and compilation of scrambling circuits), distinguishing itself from sample-and-assemble approaches. Furthermore, our construction is conceptually simple, relying only on commuting Pauli rotations that maximize circuit compression.

We note that scrambling induced by randomized Pauli rotations has also been considered in recent constructions of unitary designs~\cite{NHKW17,HLT24}. The constructions exploit, for instance, either structured (cyclically sampling certain $X$ and $Z$ exponentials)~\cite{NHKW17} or exhaustive (blindly sampling all Pauli exponentials)~\cite{HLT24} randomness to construct non-commuting rotations. While these approaches effectively generate \textit{approximate} unitary designs, we emphasize that our work achieves \textit{exact} diagonal state designs, which in turn can be built upon to synthesize unitary designs~\cite{pseudorandom}. In \cite{NHKW17} and \cite{HLT24}, the circuit depths required to achieve an approximate unitary $d$-design with an \textit{additive error} $\zeta$ are $\mathcal{O}(d[Q-\log_{2} \zeta])$ and $\mathcal{O}(d[Q-\log_{2} \zeta]\log_2 Q)$, respectively, given all-to-all device connectivity. Instead, our approach achieves an exact diagonal state $d$-design using a circuit of depth $\mathcal{O}(Q^{r - 1} \log_2 r/r!)$ with $r(d) = \lceil \log_{2}(d+1) \rceil$, as we show in \cref{sec:main results}.

Hereafter, we refer to our random states, $ \vert \chi \rangle = U \vert \chi_0 \rangle$, as \textit{quantum Hutchinson states}. To illustrate the different approaches described in this section and to motivate the quantum Hutchinson approach investigated in this work, we present an explicit comparison of classical Hutchinson, quantum Hutchinson, and one clean qubit approaches for the resolution of the identity, $\mathbb{E} \left[ \ket{\rchi} \bra{\rchi} \right] = \frac{1}{N} \mathbb{I}$, as shown in \cref{fig:QH}. It provides a visualization of how the quantum Hutchinson states achieves a high-fidelity approximation to the identity relative to the one clean qubit states with the same number of samples. Additionally, the quantum and classical Hutchinson states produce nearly identical approximations, despite the exponentially higher cost for generating the latter.

%%%% Main results %%%% 
\section{\label{sec:main results}Main results}

\subsection{\label{subsec:minimal example}Quantum Hutchinson state construction}

In this section we present the construction of our quantum Hutchinson states.

First we prepare the initial state of our circuit, which is a uniform superposition of the $Z$-computational basis elements $\Vec{n} \in \{0,1\}^{Q} $:
\begin{align}
    \ket{\rchi_0} = \ket{+}^{\otimes Q} = \frac{1}{\sqrt{N}} \sum_{\Vec{n} \in \{0,1\}^{Q} } \ket{\Vec{n}},
\end{align}
with the use of $Q=\log_2N$ Hadamard gates. 
% Here $Q$ denotes the number of qubits.

We evolve this initial state under the following diagonal Hamiltonian of spin-glass type:
\begin{align}
    G = \sum_{i \leq j}^{Q} \gaij \Ga_i \Ga_j,
    \label{eq:G_form}
\end{align}
where $\Ga_i = \frac{1-Z_i}{2}$ denotes the number operator acting on the qubit $i$, alternatively defined by $\Ga_i\ket{\Vec{n}} = n_i \ket{\Vec{n}}$, and 
$\gaij \sim \mathrm{Uniform}(0,2\pi)$ are independent and identically distributed random variables for all pairs of qubits $i \leq j$. We will evolve under $G$ for one unit of time, and this action can be viewed as a sequence of real-time evolutions by all of the $\Gamma_i \Gamma_j$ (which all commute and hence do not introduce any Trotter error) for respective durations $\gaij$. Our quantum Hutchinson states are therefore prepared as 
\begin{align}
    \ket{\rchi} = e^{ - i G} \ket{\rchi_0} = \prod_{i\leq j }^Q e^{- i \gaij \Ga_i \Ga_j} \ket{\rchi_0}, 
\label{eq:quantumhutch}
\end{align}
with $\gaij$ sampled uniformly from the interval $[0, 2\pi]$. We will verify later that this construction indeed satisfies the matching properties up to $d=3$.

% As a benchmark for comparison, we will also define the conventional Hutchinson state, 
% \begin{align}
%     \ket{\phi} = \frac{1}{\sqrt{N}} \sum_{\Vec{n}  } e^{-i \theta_{\Vec{n} } } \ket{\Vec{n}},
%       \label{eq:hutch}
% \end{align}
% where $\theta_{\Vec{n} } \sim  \mathrm{Uniform}(0,2\pi)$ are independent and identically distributed random variables for all computational basis elements  $\Vec{n}$. We note that the conventional Hutchinson construction is \emph{not tractably realizable on quantum hardware} but serves rather as an ideal standard for trace estimator variance.

\subsection{\label{subsec:aside} Construction with a single random time}

As a theoretical question, it is interesting to consider whether it is possible to construct a suitable quantum Hutchinson state as $\ket{\rchi} = e^{-i G t} \ket{\rchi_0}$, where $G$ is a single fixed Hamiltonian and $t$ is a single random time. In fact, as we show in \cref{app:random_time}, it is possible to achieve this goal where $G$ is taken to be a linear combination of the diagonal operators $\Gamma_i$ and $\Gamma_i \Gamma_j$. However, the random time $t$ must be exponentially long in the number of qubits $Q$. Ultimately it is convenient to view such a long real-time evolution as a sequence of individual real-time evolutions, whose durations are drawn from a different joint distribution than that of our main construction, according to the Hamiltonian components $\Gamma_i$ and $\Gamma_i \Gamma_j$. From this point of view, the construction \cref{eq:quantumhutch} emerges as a simpler alternative in which all the short random times are drawn independently. Moreover, the construction \cref{eq:quantumhutch} enjoys the same hardware advantages of real-time evolution under a fixed set of elementary two-qubit Hamiltonians. Nonetheless, due to potential theoretical interest, we give further detail on the single-time construction in \cref{app:random_time}, and we comment that in fact this work was initially motivated by this perspective.

\subsection{\label{subsec:stats}Statistical properties}

% Now we show that the quantum Hutchinson state construction of~\cref{eq:quantumhutch} achieves an exact stochastic resolution of identity, \textcolor{blue}{i.e., that $\mathbb{E} \left[ \ket{\rchi} \bra{\rchi} \right] = \frac{1}{N} \mathbb{I}$}. Moreover, the variance matches that of the ideal Hutchinson states furnished by~\cref{eq:hutch}. Both statements are implied by the following theorem.

Now we show that the quantum Hutchinson state constructed in~\cref{eq:quantumhutch} is a diagonal 3-design.

\bigskip

\noindent \textbf{Theorem 1.}  Consider the quantum Hutchinson and conventional Hutchinson states drawn randomly according to~\cref{eq:quantumhutch} and~\cref{eq:hutch}, respectively. The moments of these two constructions match up to third order in the sense that, for $d \leq 3$, 
\begin{align}
   \mathbb{E}\left[ \left(\ket{\rchi} \bra{\rchi}\right)^{\otimes d} \right] = \mathbb{E}\left[ \left(\ket{\phi} \bra{\phi}\right)^{\otimes d} \right].
\end{align} 

\noindent \textit{Proof.} We first observe that the Hutchinson construction admits the statistical moments 
\begin{widetext}
    \begin{align}
    \mathbb{E}\left[ \left(\ket{\phi} \bra{\phi}\right)^{\otimes 1} \right] &= \frac{1}{N} \sum_{\Vec{n}} \ket{\Vec{n}} \bra{\Vec{n}}, \label{eq:sri1}\\
     \mathbb{E}\left[ \left(\ket{\phi} \bra{\phi}\right)^{\otimes 2} \right] &= \frac{1}{N^2} \mathop{\sum_{\Vec{m}, \Vec{m}', \Vec{n}, \Vec{n}':}}_{(\Vec{m}, \Vec{m}')  = \mathcal{P}_{2}(\Vec{n}, \Vec{n}') \,  \mathrm{for\,some}\, \mathcal{P}_2 \in S_2 } \ket{\Vec{m} \Vec{m}'} \bra{\Vec{n} \Vec{n}'}, \label{eq:sri2}\\
     \mathbb{E}\left[ \left(\ket{\phi} \bra{\phi}\right)^{\otimes 3} \right] &= \frac{1}{N^3} \mathop{\sum_{\Vec{m}, \Vec{m}', \Vec{m}'', \Vec{n}, \Vec{n}', \Vec{n}'':} }_{(\Vec{m}, \Vec{m}', \Vec{m}'') = \mathcal{P}_{3}(\Vec{n}, \Vec{n}', \Vec{n}'') \,  \mathrm{for\,some}\, \mathcal{P}_3 \in S_3} \ket{\Vec{m} \Vec{m}' \Vec{m}''} \bra{\Vec{n} \Vec{n}' \Vec{n}''},
\end{align}
\end{widetext}
where the $\mathcal{P}_{d} \in S_d$ appearing in the summations above indicate permutations of $d$ bitstrings. The vector equality constraints up to a permutation arise from the statistical independence of random phases $\theta_{\Vec{n}}$ in \cref{eq:quantumhutch}.

Next we consider the quantum Hutchinson construction. Note that if $\gamma \sim \mathrm{Uniform}(0,2\pi)$, then for any integer $k$, the expectation $\mathbb{E} [e^{i k \gamma}]$ is one if $k=0$ and zero otherwise. This identity enables the computations: 

\begin{widetext}
    \begin{align}
     \mathbb{E}\left[ \left( \ket{\rchi} \bra{\rchi} \right)^{\otimes 1} \right] &= \frac{1}{N} \sum_{\Vec{m}, \Vec{n}} \ \prod_{i \leq j}^{Q} 
     \mathbb{E} \big[  e^{ i (n_i n_j - m_i m_j) \gamma } \big]
          \ket{\Vec{m}} \bra{\Vec{n}},\\
     & = \frac{1}{N} \sum_{\Vec{m}, \Vec{n}} \  \prod_{i \leq j}^{Q} \delta_{n_i n_j, m_i m_j} \ket{\Vec{m}} \bra{\Vec{n}}, \\
     &= \frac{1}{N} \mathop{\sum_{\Vec{m}, \Vec{n}:}}_{\Vec{m} \otimes \Vec{m}  = \Vec{n} \otimes \Vec{n}}  \ket{\Vec{m}} \bra{\Vec{n}} \ \stackrel{\textbf{(Lemma 1)}}{=} \  \mathbb{E}\left[ \left(\ket{ \phi } \bra{ \phi }\right)^{\otimes 1} \right], \label{eq:1st_moment} \\
     \mathbb{E}\left[ \left(\ket{\rchi} \bra{\rchi}\right)^{\otimes 2} \right] &= \frac{1}{N^2} \sum_{\Vec{m}, \Vec{m}', \Vec{n}, \Vec{n}'} \ \prod_{i \leq j}^{Q} \delta_{n_i n_j + n'_i n'_j, m_i m_j + m'_i m'_j} \ket{\Vec{m} \Vec{m}'} \bra{\Vec{n} \Vec{n}'}, \label{eq:2nd_moment_char} \\
     & = \frac{1}{N^2}  \mathop{\sum_{\Vec{m}, \Vec{m}', \Vec{n}, \Vec{n}':}}_{\Vec{m} \otimes \Vec{m} + \Vec{m}' \otimes \Vec{m}'  = \Vec{n} \otimes \Vec{n} + \Vec{n}' \otimes \Vec{n}'} \ket{\Vec{m} \Vec{m}'} \bra{\Vec{n} \Vec{n}'} \ \stackrel{\textbf{(Lemma 1)}}{=} \ \mathbb{E}\left[ \left(\ket{\phi} \bra{\phi }\right)^{\otimes 2} \right], \label{eq:2nd_moment} \\
     \mathbb{E}\left[ \left(\ket{\rchi} \bra{\rchi}\right)^{\otimes 3} \right] &= \frac{1}{N^3}  \mathop{\sum_{\Vec{m}, \Vec{m}', \Vec{m}'', \Vec{n}, \Vec{n}', \Vec{n}'':}}_{\Vec{m} \otimes \Vec{m} + \Vec{m}' \otimes \Vec{m}' + \Vec{m}'' \otimes \Vec{m}'' = \Vec{n} \otimes \Vec{n} + \Vec{n}' \otimes \Vec{n}' + \Vec{n}'' \otimes \Vec{n}''} \ket{\Vec{m} \Vec{m}' \Vec{m}''} \bra{\Vec{n} \Vec{n}' \Vec{n}''} \ \stackrel{\textbf{(Lemma 1)}}{=} \  \mathbb{E}\left[ \left(\ket{\phi} \bra{\phi }\right)^{\otimes 3} \right].\label{eq:3rd_moment}
\end{align}
\end{widetext}

To arrive at \cref{eq:1st_moment,eq:2nd_moment,eq:3rd_moment} in the calculations, we have used the following lemma: 

\bigskip

\noindent \textbf{Lemma 1.} For vectors $\Vec{m}_1, \Vec{m}_2, \Vec{m}_3, \Vec{n}_1, \Vec{n}_2, \Vec{n}_3 \in \{ 0, 1 \}^Q$ and integer $1 \leq d \leq 3$, the matrix identity, 
\begin{align}
    \sum_{i=1}^{d} \Vec{m}_i \otimes \Vec{m}_i = \sum_{i=1}^{d} \Vec{n}_i \otimes \Vec{n}_i,
\end{align}
holds if and only if $(\Vec{m}_1, \cdots, \Vec{m}_d) = \mathcal{P}_d (\Vec{n}_1, \cdots, \Vec{n}_d)$ for some permutation $\mathcal{P}_d \in S_d$. Here $\otimes$ is understood as the vector outer product so each term $\Vec{m}_i \otimes \Vec{m}_i$ can be viewed as a symmetric matrix $\Vec{m}_i^{\top} \Vec{m}_i$, or equivalently a 2-tensor, with entries $(\Vec{m}_i \otimes \Vec{m}_i)_{ab} = m_{i,a} m_{i,b}$.

\bigskip

The proof of Lemma 1 is presented within~\cref{app:prooflemma}, where we also explain why the lemma does not apply to the case $d=4$. This thus completes the proof of Theorem 1. It is worth noting that the exact same proof holds if we replace the uniform random sampling of time durations $\gamma_{ij}$ over $(0,2\pi)$ with a uniform sampling from the discrete set $\{0, \frac{\pi}{2}, \pi, \frac{3\pi}{2} \}$ as can be confirmed by direct calculation. $\square$

% \textcolor{blue}{\noindent\rule{8.25 cm}{0.3pt}}

Now let us illustrate how Theorem 1 is applied to yield guarantees on our random states $\ket{\rchi}$. First, the base case $d=1$, together with~\cref{eq:sri1}, establishes the stochastic resolution of identity, $\mathbb{E} \left[ \ket{\rchi} \bra{\rchi} \right] = \frac{1}{N} \mathbb{I}$, essential in many randomization tasks. Second, we examine the case $d=2$. Theorem 1 in this case, combined with~\cref{eq:sri2}, implies that 
\begin{align}
\begin{split}
        \mathbb{E}\left[ \left(\ket{\rchi} \bra{\rchi}\right)^{\otimes 2} \right] = \Big(  \frac{1}{N} & \sum_{\Vec{n}} \ket{\Vec{n}}  \bra{\Vec{n}} \Big)^{\otimes 2} \\
        &+ \frac{1}{N^2} \sum_{\Vec{m} \neq \Vec{n}} \ket{\Vec{m} \Vec{n}} \bra{\Vec{n} \Vec{m}},
  \end{split}
\end{align}
recovering the well-known second moment of conventional Hutchinson states:
\begin{align}
     \mathbb{E} \Big[ \abs{\braket{\rchi|A|\rchi}}^2 \Big] = \abs{\mathbb{E} \braket{\rchi|A|\rchi}}^2 + \frac{1}{N^2} \sum_{\Vec{m} \neq \Vec{n}} \abs{\braket{\Vec{m} | A |\Vec{n}}}^2,
     \label{eq:trace_var}
\end{align}
where $A \in \mathbb{C}^{N \times N}$ and $\mathbb{E} \braket{\rchi|A|\rchi} = \frac{1}{N} \Tr[A]$.

Alternatively we can express the variance as 
\begin{align}
     \mathbb{V} \left[ \braket{\rchi|A|\rchi} \right] = \frac{1}{N^2} \sum_{\Vec{m} \neq \Vec{n}} \abs{\braket{\Vec{m} | A |\Vec{n}}}^2 
     \leq \frac{1}{N^2} \Vert A \Vert_{\mathrm{F}}^2,
     \label{eq:trace_var2}
\end{align}
where we have a succinct bound in terms of the Frobenius norm of the operator $A$. Note that the exact expression for the variance can often be reduced through a similarity transformation, for example if $A$ enjoys certain symmetries~\cite{HCT_2023}. By Chebyshev's inequality, taking an empirical average over $K$ randomly sampled states, we expect that with high probability 
\begin{align}
   \abs{ \frac{1}{K} \sum_{k=1}^{K} \braket{\rchi_k|A|\rchi_k} -  {\frac{1}{N} \Tr\left[ A\right]}  } \leq \frac{1}{N} \Vert A \Vert _{\mathrm{F}} \,   \epsilon,
\end{align}
where $\epsilon = \mathcal{O}(\frac{1}{\sqrt{K}})$ independently of $A$. In this sense, we say that the fluctuations of the estimator for the normalized operator trace are on the order of $\frac{1}{N} \Vert A \Vert_{\mathrm F}$.

Finally, we explore how our simple construction can be generalized naturally to higher-order diagonal designs by increasing the interaction locality in the Hamiltonian $G$. We focus on diagonal $d$-designs with $4 \leq d \leq 7$ as an illustrative example, while demonstrating that the underlying argument extends inductively to a general diagonal $d$-design.
\bigskip

\noindent \textbf{Corollary 1.}  The quantum Hutchinson states $\ket{\rchi} = e^{ - i G} \ket{\rchi_0}$ generated by the following 3-local Hamiltonian,
\begin{align}
    G =  \sum_{i \leq j \leq k}^{Q} \gamma_{ijk} \Ga_i \Ga_j \Ga_k,
    \label{eq:G_4design}
\end{align}
form a diagonal 7-design, 
% \ML{don't you mean 7-design? Check for other errors of this type in the doc.}
provided that all the durations $\gamma_{ijk} \sim \mathrm{Uniform}(0,2\pi)$ or $\gamma_{ijk} \sim \mathrm{Uniform}\{\frac{\pi}{4}f: f \in \mathbb{Z}_{8} \}$ are drawn independently. More generally, an $r$-local Ising Hamiltonian of the form, 
\begin{align}
    G =  \sum_{i_1 \leq i_2 \leq \cdots \leq i_r}^{Q} \gamma_{i_1 i_2 \ldots i_r} \Ga_{i_1} \Ga_{i_2} \cdots \Ga_{i_r},
\end{align}
generates a diagonal $d$-design of order up to $d=2^r - 1$, provided that $\gamma_{i_{1} i_{2} \cdots i_{r}} \sim \mathrm{Uniform}(0,2\pi)$ or $\gamma_{i_{1} i_{2} \cdots i_{r}} \sim \mathrm{Uniform}\{\frac{2\pi}{d+1}f: f \in \mathbb{Z}_{d+1} \}$ are drawn independently.
\bigskip

The proof of Corollary 1 follows from Theorem 1 and is outlined in~\cref{app:prooflemma}. $\square$

\subsection{Reconfigurability on digital platforms}
Next we examine the hardware requirements for our diagonal $3$-design on a digital quantum platform. In particular, all quantum Hutchinson instances can be realized with a single reconfigurable quantum circuit.

Observe that by using $\Gamma_i \Gamma_j = \frac{1}{4} (1 - Z_i - Z_j + Z_i Z_j)$, the Hamiltonian $G$ from~\cref{eq:G_form} can be rewritten, up to a global phase, in terms of the Pauli $Z$ operators as the Ising-type Hamiltonian,
\begin{align}
    G = \sum_{i < j}^{Q} h_{ij} Z_i Z_j + \sum_{i=1}^{Q} h_i Z_i,
\end{align}
with the Ising couplings and external field strengths given by $h_{ij} = \frac{\gaij}{4}$ and $h_i = -\frac{\gamma_{ii}}{2} - \sum_{j > i} \frac{\gaij}{4}$. 
The unitary time evolution in the $Z$-basis is hence a product of elementary diagonal gates,
\begin{align}
    e^{-iG t} &= \left( \prod_{i\neq j}^{Q} e^{-ih_{ij} Z_i Z_j t} \right) \left( \prod_{i=1}^{Q} e^{-ih_{i} Z_i t} \right), \\
    & = \left( \prod_{i < j}^{Q} R_{zz}(4 h_{ij}t|i,j) \right)   \left( \bigotimes_{i=1}^{Q} R_{z}(2h_i t|i) \right),
\end{align}
where $R_z(\phi|i)$ and $R_{zz}(\phi|i,j)$ are one- and two-qubit $Z$-rotations of angle $\phi$ respectively. Each two-qubit rotation can be further unfolded as a CNOT-conjugated one-qubit rotation,
\begin{align}
    R_{zz}(\phi|i,j) = C_x(i,j)  R_{z}(\phi|j) C_x(i,j), 
    % \left( {\rm CNOT}\vert_{j}^{i} \right)
\end{align}
where the CNOT, $C_x(i,j)$, acts on the control and target qubits $(i,j)$. Hence, the circuit can be efficiently synthesized using only one-qubit $Z$-rotations and two-qubit CNOT gates.

Notably, the quantum Hutchinson state sampling only involves modification of the $\mathcal{O}(Q^2)$ rotation angles but does not involve replacement or reordering of any gates. Moreover, by exploiting gate commutation relations, we can bound both the gate count and circuit depth required to produce a quantum Hutchinson state (cf. Theorem 2 below). The circuit we have presented yields a na\"{i}ve  gate count scaling of $\mathcal{O}(Q^2)$, with $\frac{Q(Q + 1)}{2}$ $R_z$ gates and $Q(Q-1)$ CNOT gates. While the implementation of this circuit is indeed feasible on devices with native all-to-all connectivity, we can slightly 
%improve the na\"{i}ve gate count 
reduce the CNOT gate count
and moreover show that the circuit depth is $\mathcal{O}(Q)$.

\bigskip
\noindent \textbf{Theorem 2.} $e^{-i G}$ can be implemented using a single circuit with depth no greater than $D = 9Q-2$, where the circuit contains $\frac{Q(Q+1)}{2}$ $R_z$ gates and at most $N_{\rm CNOT} = \lfloor \frac{5Q^2 - 3Q-2}{6} \rfloor$ CNOT gates.

\bigskip

\begin{figure}[t!]
    \begin{tikzpicture}
    \newcommand{\numnodes}{8}
    \foreach \ii in {1, ..., \numnodes}{   \pgfmathsetmacro{\angle}{90 -      (\ii-1)*360/\numnodes}
        \node[circle, scale=0.75, draw, fill=green!40] (x\ii) at (\angle:2cm) {\ii};
    }
    \foreach \ii in {1, ..., \numnodes}{   \pgfmathsetmacro{\prev} 
        {int(ifthenelse(\ii==1, \numnodes, \ii-1))}
        \pgfmathsetmacro{\next}{int(ifthenelse(\ii==\numnodes, 1, \ii+1))}
        \draw[thin, dashed] (x\ii) -- (0,0) (x\prev) -- (x\ii) -- (x\next);
        \foreach \jj in {\ii, ..., \numnodes}{
        \pgfmathsetmacro{\drawoptions}{ifthenelse(
                \jj==(\ii+\numnodes/2) || \jj==\prev || \jj==\next,
                "none", "gray")}
            \path[draw=\drawoptions, thin, dashed] (x\ii) -- (x\jj);
        }  
    }
    \draw[thick,cyan] (x8) -- (x1) -- (x2) -- (x8);
    \draw[thick,cyan] (x2) -- (x3) -- (x4) -- (x2);
    \draw[thick,cyan] (x4) -- (x5) -- (x6) -- (x4);
    \draw[thick,cyan] (x6) -- (x7) -- (x8) -- (x6);

    \draw[thick,red] (x7) -- (x1) -- (x3) -- (x7);
    \draw[thick,blue] (x7) -- (x2) -- (x5) -- (x7);
    \draw[thick,blue] (x8) -- (x3) -- (x5) -- (x8);
    \end{tikzpicture}
    \caption{Edge-disjoint triangles in a complete graph with $|\mathcal{V}|=Q=8$ vertices. The colors distinguish triangles with different connectivities.}
    \label{fig:triang_graph}
\end{figure}
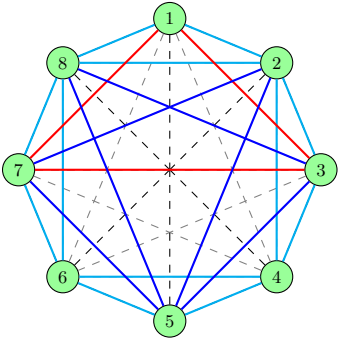

\noindent \textit{Proof.} We exploit the commutation or rewriting rules for $R_z$ and CNOT gates,
\begin{align}
    C_x(j,k)R_{zz}(\phi|i,j) = R_{zz}(\phi|i,j) C_x(j,k), \\
    C_x(i,j) C_x(j,k) C_x(i,k) = C_x(j,k) C_x(i,j),
\end{align}
where $(i,j,k)$ indexes a triplet of distinct qubit registries. When combined, the rules help improve local compilation as illustrated graphically below,

\begin{center}
  \begin{quantikz}[row sep={5mm,between origins}, column sep={2mm}]
  &\ctrl{2} &\qw &\ctrl{2}              &\ctrl{1} &\qw  &\ctrl{1} &\qw &\qw &\qw &\qw\\
  &\qw &\qw  &\qw &\targ{} &\gate{R_z}  &\targ{} &\ctrl{1} &\qw &\ctrl{1} &\qw\\
  &\targ{} &\gate{R_z} &\targ{} &\qw &\qw &\qw &\targ{} &\gate{R_z} &\targ{} &\qw
  \end{quantikz}
  = 
  \begin{quantikz}[row sep={5mm,between origins}, column sep={2mm}]
  &\ctrl{2} &\ctrl{1} &\qw              &\qw &\qw &\ctrl{1} &\qw &\qw\\
  &\qw &\targ{} &\qw &\ctrl{1} &\gate{R_z} &\targ{} &\ctrl{1} &\qw\\
  &\targ{} &\gate{R_z} &\qw &\targ{} &\qw &\gate{R_z} &\targ{} & \qw
  \end{quantikz},
\end{center} 

\noindent resulting in a reduction in both the gate count and circuit depth for a three-qubit block. Recall that each $R_{zz}(\phi|i,j)$ uniquely captures a two-body interaction $h_{ij}Z_i Z_j$ in $G$. So the overall resource reduction from our improved local compilation can be estimated by identifying the triplets or triangles  $(i,j,k)$ that lack shared edges in an all-to-all graph representing our Ising Hamiltonian. This is a topic deeply examined in graph theory and combinatorics~\cite{SPENCER19681,assmus_key_1992,feder2012packing}. In particular, we highlight the fact that in a complete graph, $\mathcal{G} = (\mathcal{V}, \mathcal{E})$ with $\abs{\mathcal{E}} = \frac{|\mathcal{V}|(|\mathcal{V}|-1)}{2}$, the maximal number of such edge-disjoint triangles is at least $N_{a} \geq \frac{(|\mathcal{V}|-1)(|\mathcal{V}|-2)}{6}$. For example, \cref{fig:triang_graph} shows the assembly of different three-qubit blocks in a $8$-qubit complete graph, where a set of $7$ edge-disjoint triangles is highlighted. The number of `dangling' edges untouched by the triangles is therefore bounded by $N_{b} \leq \abs{\mathcal{E}} - 3N_{a} = Q - 1$, which gives a total CNOT count of
\begin{align}
    N_{\rm CNOT} \leq 5 N_{a} + 2 N_{b}.
\end{align}
Moreover, it is apparent that any two triangles commute with each other unless they share a common vertex. This observation allows us to bound the circuit depth, 
\begin{align}
    D \leq 12 \lfloor{\frac{Q}{2}}\rfloor + 3 N_b + 1,
    \label{eq:twobody_depth}
\end{align}
where the first two terms in \cref{eq:twobody_depth} account for the $\lfloor{\frac{Q}{2}}\rfloor$ layers of triangles with varying connectivities, as marked by the different colors in \cref{fig:triang_graph}, and $N_b$ layers of untouched dangling edges. We also include the layer of $R_z$ gates that capture the external fields $\{h_i Z_i\}_{i=1}^{Q}$ in $G$.

(For the special cases $Q \equiv 1\mod 6$ or $Q \equiv 3\mod 6$, we saturate an optimal number of edge-disjoint triangles, $N_a = \frac{Q(Q-1)}{6}$. This therefore implies a tighter bound of $N_{\rm CNOT} \leq \frac{5Q^2 - 5Q}{6} $ and $D \leq 6Q+1$ in these cases.) $\square$

\bigskip

For higher-order diagonal designs generated via $r$-local Hamiltonian, a similar analysis allows us to establish that the scrambling circuit has a depth of $\mathcal{O}(Q^{r-1}/r!)$, where we recall $r = \lceil \log_2 (d + 1) \rceil$ as indicated in \cref{subsec:stats}.

\subsection{Reconfigurability on analog platforms}

In this section, we proceed to investigate the hardware implementation of our diagonal $3$-design on a dynamics-driven analog platform, where the construction based on the real-time evolution yields significant advantages. Our analysis primarily centers around systems of interacting neutral atoms undergoing Rydberg excitations~\cite{Bernien2017,Bluvstein2022}. Although our focus is specific, similar considerations can hold for other analog models.

Recall that a general Rydberg Hamiltonian takes the form
\begin{align}
    G_{\rm Rydberg} =  \sum_{i=1}^{Q} \Omega_i X_i +  \sum_{i=1}^{Q} \Delta_i \Ga_i + \sum_{i<j}^{Q} V_{ij} \Ga_i \Ga_j,
    \label{eq:Rdy_Ham}
\end{align}
where $\Omega_i$ and $\Delta_i$ determine the local Rabi frequency and detuning, respectively, on the atomic site $i$, while $V_{ij} \propto \frac{1}{r_{ij}^6}$ sets the repulsive van der Waals interaction between sites $i$ and $j$ separated by a distance of $r_{ij}$. Therefore our spin-glass Hamiltonian $G$ is natively Rydberg with vanishing Rabi driving, $\Omega_i \equiv 0$.

In addition, preparation of the initial state can be accomplished by unit-time evolution under another Rydberg Hamiltonian of the form of~\cref{eq:Rdy_Ham}, where $\Omega_i = - \frac{\Delta_i}{2} = -\frac{\pi}{2\sqrt{2}}$ and $V_{ij} \equiv 0$. This is the case because
\begin{align}
    \ket{\rchi_0} = \bigotimes_{i=1}^{Q} e^{i\frac{\pi}{2\sqrt{2}}(X_i + Z_i)} \ket{\Vec{0}},
\end{align}
where physical ground state $\ket{\Vec{0}}$ is the native atomic configuration.

The overall evolution that we require, including the preparation of $\ket{\rchi_0}$ can therefore be
produced using the time-dependent waveforms, 
\begin{align}
    \begin{cases}
        (\Omega_i, \Delta_i) = (-\frac{\pi}{2\sqrt{2}}, \frac{\pi}{\sqrt{2}}), &~  0 \leq t \leq 1 \\
        (\Omega_i, \Delta_i) = (0, \gamma_{ii}), &~ 1 \leq t \leq 2
    \end{cases},
\end{align}
prescribing piecewise-constant local Rabi drivings as well as detunings.

Moving forward, we discuss the simulation of the remaining pairwise coupling terms with physical interactions $V_{ij}$. Na\"{i}vely, given positive weights $\gamma_{ij} \overset{\mathrm{iid}}{\sim} \mathrm{Uniform}(0, 2\pi)$, we must choose the $Q$ atomic positions $ \Vec{r}_i$ so that the resulting distances $r_{ij} = \lVert \Vec{r}_i - \Vec{r}_j \rVert_2$ satisfy $V_{ij}(r_{ij}) = \gaij$. However, this embedding problem is NP-hard when $Q$ exceeds the spatial dimension (which is $3$ at most), and a solution does not exist in most cases.

Instead we formulate an alternative approach by exploiting the commutation of the diagonal operators $\Gamma_i$. As a first attempt we consider the decomposition $e^{-iGt} =  \prod_{i<j} e^{- i G_{ij} t}$ where
\begin{align}
     G_{ij} := \gaij \Ga_i \Ga_j + \frac{1}{Q-1} (\gamma_{ii} \Ga_i + \gamma_{jj} \Ga_j),
\end{align}
which holds because $\Gamma_i = \Gamma_i \Gamma_i$ for each $i$. The regrouping of the Hamiltonian contributions suggests that we should achieve the analog evolution $e^{-i G t}$ using $\frac{Q(Q-1)}{2}$ analog blocks $e^{-iG_{ij}t}$ each of which couples only a single pair of atoms, and between which atom positions are dynamically shuffled.

However, by considering a more sophisticated coupling within each analog block we can bring the block depth down to $Q$. Theorem 3 below states this result and moreover quantifies the atom shuffling that we require between blocks in terms of the number of SWAP gates, each of which is realized as the physical swapping of a pair of atoms. In this construction, it is more convenient to work with simplified $\gamma_{ij}$ that are sampled i.i.d. from the uniform distribution over the discrete set $\big\{0, \frac{2\pi}{3}, \frac{4\pi}{3} \big\}$. If $\gamma$ is so distributed, then for any integer $k$, the expectation $\mathbb{E} [e^{i k \gamma}]$ is one if $k$ is an integer multiple of $3$ and zero otherwise. Therefore the variance equality from Theorem 1 holds just as well in this case. 

\begin{figure}[t!]
    \begin{tikzpicture}
    \newcommand{\numnodes}{7}
    \foreach \ii in {1, ..., \numnodes}{   \pgfmathsetmacro{\angle}{90 -      (\ii-1)*360/\numnodes}
        \node[circle, scale=0.75, draw, fill=green!40] (x\ii) at (\angle:2cm) {\ii};
    }
    \node[circle, scale=0.75, draw, fill=green!40] (x8) at (0:0cm) {8};
    
    \foreach \ii in {1, ..., \numnodes}{ 
    \draw[thin, dashed] (x8) --(x\ii);
    }
    
    \draw[thick, gray] (x7) -- (x1) -- (x2) -- (x7);
    \draw[thick, gray] (x2) -- (x3) -- (x4) -- (x2);
    \draw[thick, gray] (x4) -- (x5) -- (x6) -- (x4);
    \draw[thick, gray] (x6) -- (x7) -- (x7) -- (x6);
    \draw[thick, gray] (x7) -- (x1) -- (x6) -- (x7);
    \draw[thick, gray] (x7) -- (x3) -- (x4) -- (x7);
    \draw[thick, gray] (x1) -- (x5) -- (x4) -- (x1);
    \draw[thick, gray] (x6) -- (x2) -- (x3) -- (x6);
    \draw[thick, gray] (x7) -- (x1) -- (x3) -- (x7);
    \draw[thick, gray] (x7) -- (x2) -- (x5) -- (x7);
    \draw[thick, gray] (x7) -- (x3) -- (x5) -- (x7);

    \draw[thick, dashed, blue] (x8) -- (x1);
    \draw[thick, blue] (x2) -- (x7);
    \draw[thick, blue] (x3) -- (x6);
    \draw[thick, blue] (x4) -- (x5);

    \draw[thick, dashed, red] (x8) -- (x2);
    \draw[thick, red] (x1) -- (x3);
    \draw[thick, red] (x4) -- (x7);
    \draw[thick, red] (x6) -- (x5);
    
    \end{tikzpicture}
    \caption{Edge-coloring of a complete graph with $|\mathcal{V}|=Q=8$ vertices. The colors partition all the edges into $(Q-1)$ groups. Each group represents a unique atom configuration consisting of disjoint doublets. For purpose of visualization, we highlight two of the colored groups (red and blue respectively).}
    \label{fig:coloring_graph}
\end{figure}
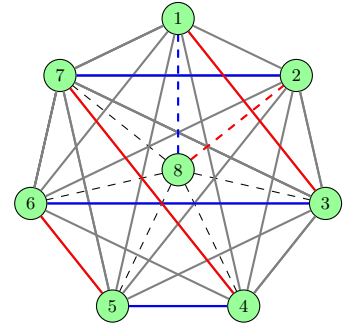

\bigskip
\noindent \textbf{Theorem 3.} $e^{-i G}$ can be implemented on a Rydberg simulator with at most $Q$ analog blocks. Each of these analog blocks performs Rydberg evolution of time $\mathcal{O}(1)$ for a one-dimensional chain of $Q$ Rydberg atoms. The circuit also involves a total of at most $N_{\rm SWAP} = \frac{Q(Q-1)}{2}$ SWAP gates, i.e., the physical swaps of two atoms, which appear in between the analog blocks.

\bigskip

\noindent \textit{Proof.} The use of the discrete couplings $\gamma_{ij}$ allows us to restrict to only three physical distances $\Lambda_2 < \Lambda_1 < \Lambda_{0}$ such that $V_{ij}(\Lambda_{2}) = \frac{4\pi}{3}$, $V_{ij}(\Lambda_{1}) = \frac{2\pi}{3}$, and $V_{ij}(\Lambda_{0}) = 0$. This choice is possible due to the rapid decay of the van der Waals interaction.

We now invoke classic edge coloring results for a complete graph of $Q$ vertices corresponding to the $Q$ qubits. Specifically, the edges of this graph can be decorated with at most $Q$ different colors so that no incident edges have the same color. In general, edge colorings can be assigned efficiently with $\mathcal{O}({\rm poly}(Q))$ runtime~\cite{baranyai1974factrization,beineke_wilson_2015}, but we make use of a very specific construction for the complete graph.

Without loss of generality, we assume that $Q$ is even and use the following coloring scheme. We distribute $Q-1$ vertices evenly on a circle to create a regular polygon and locate the remaining vertex at the center of the circle, as illustrated in \cref{fig:coloring_graph}. Then we assign a distinct color to each edge connecting a central-exterior vertex pair (as indicated by dashed lines in \cref{fig:coloring_graph}), and color the edges perpendicular to the central edge (as solid lines in \cref{fig:coloring_graph}). This assignment uses $\frac{|\mathcal{E}|}{|\mathcal{V}|/2} = Q-1$ colors to cover our complete graph. 

\begin{figure}[t!]
    \centering
    \begin{tikzpicture}
    \tikzset{>=stealth}

    \node[below] at (0.25,-0.25) {$\Lambda_2$};
    \node[below] at (1,-0.25) {$\Lambda_0$};   
    \node[below] at (4.7,-0.25) {$\Lambda_1$};   

    %%%%%%%%%%% red color %%%%%%%%%%%
    %%%% 1st row
    \node[circle, very thick, scale=0.8, draw=red, fill=green!40] (xr1) at (0,2) {3};
    \node[circle, very thick, scale=0.8, draw=red, fill=green!40] (xr2) at (0.5,2) {1};

    \node[circle, very thick, scale=0.8, draw=red, fill=green!40] (xr3) at (1.5,2) {4};
    \node[circle, very thick, scale=0.8, draw=red, fill=green!40] (xr4) at (2,2) {7};
     
    \node[circle, very thick, scale=0.8, draw=red, fill=green!40] (xr5) at (3,2) {5};
    \node[circle, very thick, scale=0.8, draw=red, fill=green!40] (xr6) at (3.5,2) {6};
    
    \node[circle, very thick, scale=0.8, draw=red, fill=green!40] (xr7) at (4.4,2) {2};
    \node[circle, very thick, scale=0.8, draw=red, fill=green!40] (xr8) at (5,2) {8};
    
    %%%% 2nd row 
    \node[circle, very thick, scale=0.8, draw=red, fill=green!40] (xr1_) at (0,1) {3};
    \node[circle, very thick, scale=0.8, draw=red, fill=green!40] (xr2_) at (0.5,1) {2};

    \node[circle, very thick, scale=0.8, draw=red, fill=green!40] (xr3_) at (1.5,1) {6};
    \node[circle, very thick, scale=0.8, draw=red, fill=green!40] (xr4_) at (2,1) {7};
     
    \node[circle, very thick, scale=0.8, draw=red, fill=green!40] (xr5_) at (3,1) {5};
    \node[circle, very thick, scale=0.8, draw=red, fill=green!40] (xr6_) at (3.5,1) {4};
    
    \node[circle, very thick, scale=0.8, draw=red, fill=green!40] (xr7_) at (4.4,1) {1};
    \node[circle, very thick, scale=0.8, draw=red, fill=green!40] (xr8_) at (5,1) {8};
    
    %%%%%%%%%%% blue color %%%%%%%%%%%
    \node[circle, very thick, scale=0.8, draw=blue, fill=green!40] (xb1) at (0,0) {3};
    \node[circle, very thick, scale=0.8, draw=blue, fill=green!40] (xb2) at (0.5,0) {6};
    
    \node[circle, very thick, scale=0.8, draw=blue, fill=green!40] (xb4) at (1.5,0) {2};
    \node[circle, very thick, scale=0.8, draw=blue, fill=green!40] (xb5) at (2,0) {7};
     
    \node[circle, very thick, scale=0.8, draw=blue, fill=green!40] (xb5) at (3,0) {5};
    \node[circle, very thick, scale=0.8, draw=blue, fill=green!40] (xb6) at (3.5,0) {4};
    
    \node[circle, very thick, scale=0.8, draw=blue, fill=green!40] (xb7) at (4.4,0) {1};
    \node[circle, very thick, scale=0.8, draw=blue, fill=green!40] (xb8) at (5,0) {8};

    \draw[ultra thick, <->] (xr2) to [bend left=50] node[above,font=\scriptsize] {1 $\leftrightarrow$ 2} (xr7);
    \draw[very thick, <->] (xr3) to [bend left=40] node[above,font=\scriptsize] {4 $\leftrightarrow$ 6} (xr6);
    \draw[very thick, <->] (xr2_) to [bend left=50] node[above,font=\scriptsize] {2 $\leftrightarrow$ 6} (xr3_);

    \shade[top color=red, bottom color=blue, shading angle=0] (5.875, 0.1) rectangle ++(0.25, 1.9);
    \draw[ultra thick, ->, draw=blue, line width=4.5pt] (6, 0.025) -- ++(0,-5pt);

    \begin{scope}[on background layer]
        \draw[very thick, gray, dotted] (-0.5,2) -- (5.5,2);
        \draw[very thick, gray, dotted] (-0.5,1) -- (5.5,1);
        \draw[very thick, gray, dotted] (-0.5,0) -- (5.5,0);
    \end{scope}
    \end{tikzpicture}
    \caption{Embedding of the edge-colored graph onto a 1D chain of interacting neutral atoms with $|\mathcal{V}| = Q = 8$ qubits. Each color, assigned on a subset of $Q/2$ edges, designates a dimer pairing of the vertices, which is reconfigured after each analog block $e^{-iG_k }$. The swapping involved in one such dimer reconfiguration (red to blue as in \cref{fig:coloring_graph}) is illustrated from top to bottom. The black bent arrows indicate the sequence of vertex transpositions needed to transition from the old to the new color assignment.}
    \label{fig:1D lattice}
\end{figure}
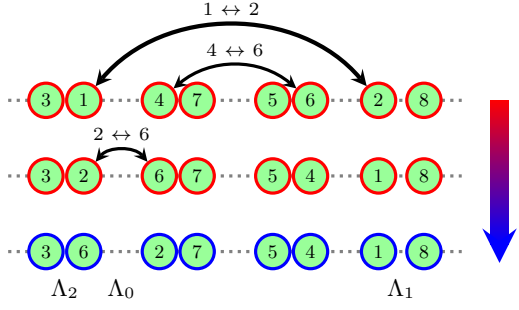

Let $\{\mathcal{E}_i \}_{i=1}^{Q-1}$ be a partition of the edge set $\mathcal{E}$,
\begin{align}
    \bigsqcup_{i=1}^{Q-1} \mathcal{E}_i = \mathcal{E},
\end{align}
resulting from the coloring scheme above. Our analog evolution can thus be achieved using $Q-1$ blocks,
\begin{align}
    e^{-iG } = \prod_{k=1}^{Q-1} e^{-iG_{k} } =  \prod_{k=1}^{Q-1} \prod_{(i,j) \in \mathcal{E}_{k} } e^{- i G_{ij} },
\end{align}
where the Hamiltonian $G_k$ for each analog block only includes interactions between atom dimers that share the same color assignment. The block Hamiltonian $G_k$ can be embedded onto a one-dimensional lattice with variable lattice spacing $\Lambda_0$, $\Lambda_1$ and $\Lambda_2$ as sketched in \cref{fig:1D lattice}. To change the dimer configuration from $G_k$ to $G_{k+1}$, we permute $Q/2$ vertices on the lattice since we can redistribute them to pair with the remaining half of the vertices. For example, the pairing $(3,1) \mapsto (3,6)$ in  \cref{fig:1D lattice} from red to blue leaves the vertex $3$ untouched. Such a permutation can canonically be written as a sequence of at most $\frac{Q}{2} - 1$ elementary transpositions. A transposition of the two vertices $i$ and $j$ corresponds to the physical swapping of two atom positions. Such atom swaps upon the application of each analog block are shown schematically in \cref{fig:1D lattice}. We notice that each set of $\frac{Q}{2}$ distinctly colored edges can be paired with one of the exterior edges $(1,2), (2,3),\ldots,$ etc. Since our graph of polygon exhibits a discrete rotational symmetry along its exterior edges, all the $\abs{\mathcal{V}}$ edges can be decorated as we cycle through the exterior edges. Accordingly, we employ a total of $Q-1$ different permutations throughout the whole analog evolution.

Finally, we remark that the complementary case of odd $Q$ can be adapted readily from the even case, if we introduce an additional, auxiliary vertex to the graph. Combining the two cases, we thereby need at most $N_{\rm SWAP} = Q\big( \frac{Q+1}{2}-1 \big) = \frac{Q(Q-1)}{2}$ to bridge the $Q$ analog blocks.~$\square$

\bigskip

Meanwhile, analog devices with local qubit connectivities, commonly a nearest-neighbor topology, are also capable of simulating $e^{-i G}$ with a similar resource cost. For nearest-neighbor devices, the identification of the graph Hamiltonian paths \cite{hpath_cycle,hpath_graph}, for example, enables us to decompose the evolution into $\mathcal{O}(Q)$ analog blocks~\cite{hpath_quantum}.

%%%% Conclusion %%%% 
\section{Conclusion}
\label{sec:conclusion}
 
In this work, we introduce a real-time diagonal design that is efficiently implementable on near-term hardware. Leveraging time evolution under Ising-type Hamiltonians that can be conveniently realized on digital and analog quantum platforms, we construct randomized `quantum Hutchinson' states. We show that our quantum Hutchinson states match the moments that would be achieved by conventional Hutchinson states in an exponentially high-dimensional space, though the latter can not be efficiently realized with quantum resources. Furthermore, we discuss how to implement the real-time states using a single reconfigurable circuit, whose compilation across different stochastic runs remains unchanged up to the durations of the real-time evolutions. The hardware-adapted implementation suits state-of-the-art digital and analog platforms.

%%%% Acknowledgments %%%% 
\section{\label{sec:acknowledgement}Acknowledgments}
This work was funded by the U.S. Department of Energy (DOE) under Contract No.~DE-AC0205CH11231, through the Office of Science, Office of Advanced Scientific Computing Research (ASCR) Exploratory Research for Extreme-Scale Science (YS, KK, DC, RVB). This research used resources of the National Energy Research Scientific Computing Center (NERSC), a U.S. Department of Energy Office of Science User Facility located at Lawrence Berkeley National Laboratory, operated under Contract No. DE-AC02-05CH11231. ER acknowledges support from the Center for Computational Study of Excited-State Phenomena in Energy Materials (C2SEPEM) at the Lawrence Berkeley National Laboratory, which is funded by the U.S. Department of Energy, Office of Science, Basic Energy Sciences, Materials Sciences and Engineering Division, under Contract No. DE-AC02-05CH11231, as part of the Computational Materials Sciences Program. 
 NMT is grateful for support from NASA Ames Research Center.  This material is based upon work supported by the
U.S. Department of Energy, Office of Science, National
Quantum Information Science Research Centers, Superconducting Quantum Materials and Systems Center
(SQMS) under the contract No. DE-AC02-07CH11359
% NMT acknowledge funding from the NASA ARMD Transformational Tools and Technology (TTT) Project. 
and by the U.S. Department of Energy, Office of Science, Accelerated Research in Quantum Computing Centers, Quantum Utility through Advanced Computational Quantum Algorithms, grant no. DE-SC0025572. ML also acknowledges support from a Sloan Research Fellowship.

\appendix
\section*{Appendix}
\section{Construction of diagonal design with a single random time}
\label{app:random_time}

Here we detail the construction, indicated in~\cref{subsec:aside}, achieving a suitable quantum Hutchinson state with a single random time and a fixed Hamiltonian $G$.

First, for arbitrary $G$, consider the full spectral decomposition
\begin{align}
    G = \sum_{n=0}^{N-1} \lambda_n \ket{n^G} \bra{n^G},
\end{align}
with eigenpairs $(\lambda_n, \ket{n^G})_{n=0}^{N-1}$.

Let us examine the dynamics on an initial state which is a uniform superposition in the $G$-basis, namely $\ket{\chi^G} := \frac{1}{\sqrt{N}} \sum_{n=0}^{N-1} \ket{n^G}$. For a time $t \in \mathbb{R}$, we denote the corresponding time-evolved state by $\ket{\rchi(t)} = e^{-i G t} \ket{\rchi^G}$. We will view $\ket{\rchi(t)}$ as a random state by randomly sampling \emph{only} the time $t$ from a suitable scalar distribution with probability density $p(t)$. Taking an expectation $\mathbb{E}$ over the time randomness, we arrive at the operator identity,
\begin{align}
     \mathbb{E} \left[ \ket{\rchi (t) } \bra{\rchi (t) } \right] &=  \frac{1}{N} \sum_{n,m = 0}^{N-1} \mathbb{E} \big[ e^{i (\lambda_n - \lambda_m) t} \big] \ket{m^G} \bra{n^G},\\
    & = \frac{1}{N} \left( \mathbb{I} + \mathbb{B} \right),  
\end{align}
where the operator $\mathbb{B}$ is a bias in the resolution of identity, which we will show can be made zero or extremely small. (Note that in our main construction of~\cref{eq:quantumhutch}, the resolution of identity is always exactly unbiased.)

 In the $G$-basis, $\mathbb{B}$ takes the form, 
\begin{align}
     \mathbb{B} = \sum_{n \neq m}^{N-1} \varphi_t(\Dmn)  \ket{m^G} \bra{n^G},
     \label{eq:residual}
\end{align}
where $\Dmn = \lambda_n - \lambda_m$ denote the $G$-eigenvalue differences and $\varphi_t(s) := \mathbb{E} [ e^{i s t} ]$ denotes the characteristic function of the distribution $p(t)$.

A suitable choice of $p(t)$ can effectively control the bias. For example, a Gaussian distribution $p(t) = \frac{
1}{\sqrt{2\pi \sigma^2}} e^{-t^2/2\sigma^2}$ yields $\varphi_t(\Dmn) = e^{- \Dmn^2 \sigma^2/2}$. As $\sigma$ increases, the bias gets exponentially suppressed. Alternatively, the bias from the fat-tailed distribution $p(t) = \frac{ 1 }{\pi t^2/\sigma} \sin^2 (t/\sigma)$ completely vanishes for large $\sigma$. In either case, we observe from \cref{eq:residual} that a nondegenerate $G$-spectrum, \textit{i.e}, in which $\Dmn \neq 0$ for all $m \neq n$, is essential to approximately fulfill the stochastic resolution of identity bias, since indeed $\varphi_t(0) = 1$ always.

In addition to guaranteeing the stochastic resolution of identity, we also want our Hutchinson-type estimator to have an ideal variance. Then for a given operator $A$ whose trace we seek, we can compute the variance of our estimator as:
\begin{align}
   \mathbb{V}[ \bra{\chi(t)} A \ket{\chi(t)} ] =  \sum_{m \neq n} \sum_{m' \neq n'} A_{mn}^{\ast} A_{m'n'} \, \varphi_t(\Dmnp^{mn})
\end{align}
where again the expectation is taken with respect to the distribution $p(t)$ for $t$ and $A_{mn}$ are the matrix elements of $A$ in the $G$-basis. This variance will match the idealized Hutchinson variance,
\begin{align}
   \sum_{m\neq n } \vert A_{mn} \vert^2,
\end{align}
if all the nontrivial differences of eigenvalue differences, denoted $\Dmnp^{mn} = \Dmn - \Dmnp$, where $(m,n) \neq (m',n')$ and $(m,m') \neq (n,n')$, are nonzero and the characteristic function is suitably constructed as outlined above.

To make things more concrete, we consider $G$ which is diagonal in the $Z$-basis, so $\ket{\rchi_0} = \ket{\rchi^{G}} = \ket{+}^{\otimes Q}$. As a first attempt, we use the operator $G_1$ with an equispaced spectrum $\lambda_n = n$, where $n=0,\ldots, N-1$. In fact $G_1$ can be tractably constructed on the quantum computer with $Q = \log_2 N$ qubits in terms of the one-qubit number operators as 
\begin{align}
    G_1 = \sum_{k=1}^Q  2^{k-1} \Gamma_k.
\end{align}
Although the spectrum of $G_1$ is nondegenerate, the differences of eigenvalue differences are highly degenerate, hence the variance of the resulting Hutchinson-type estimator is severely suboptimal.

In order to recover performance of the classic Hutchinson estimator, we are motivated to ask: can we find a simple parametric form of the $G$-spectrum $\{\lambda_n\}$ such that the resulting `excitations' $\{ \Dmn \}_{m\neq n}$, are all distinct from one another? This same question motivates the theory of Golomb rulers in number theory~\cite{Sidon1932,golomb1972number,drakakis2009review}.

The simplest explicit specification of a Golomb ruler is the sequence,
\begin{align}
    \lambda_n = N n^2 + n, ~~ 0 \leq n \leq N-1,
\end{align}
where $\min_{m\neq n} \abs{\Dmn} = \Delta_{01} = N+1$.
It is convenient to normalize our sequence of $G$-eigenvalues, $\lambda_n = \frac{N}{N+1} n^2 + \frac{1}{N+1} n$, so that the minimal excitation is one: $\min_{m\neq n} \abs{\Dmn} = 1$.

We can construct a Hamiltonian $G_2$ achieving this spectrum in terms of $G_1$:
\begin{align}
    G_2 &= \frac{N}{N+1} G_1^2 + \frac{1}{N+1} G_1, \\
    &= \sum_{i \neq j}^{Q} h_{ij} Z_i Z_j + \sum_{i=1}^{Q} h_i Z_i + \rm{const.},
\end{align}
for suitably defined Ising-type interactions $h_{ij}$ and field strengths $h_i$. 

For an appropriate choice of time randomness, \textit{e.g.}, $p(t) = \frac{ 1 }{\pi t^2/\sigma} \sin^2 (t/\sigma)$ with $\sigma >  1$, the exact variance of the conventional Hutchinson estimator is therefore recovered.

Note, however, that the parameters $h_{ij}$ and $h_i$ are exponentially large in the number $Q$ of qubits. In order to construct gates implementing time evolution by $G$ of duration $t$ for a random time $t \sim p(t)$, we would need to compute $t h_{ij}$ and $t h_i$ modulo $2 \pi$ to a high precision. This is in principle possible but would require extended-precision arithmetic as the number of qubits grows. Since time evolution by $G$ would best be achieved by running separate time evolutions by the component terms of the Hamiltonian (whose durations are only significant modulo $2 \pi$), we recover an approach similar to our main construction~\cref{eq:quantumhutch}. Although the simplicity of \cref{eq:quantumhutch} is to be preferred, it is interesting to observe the connection.

\section{Proof of Lemma and Corollary}
\label{app:prooflemma}
\subsection{Proof of Lemma 1}
\label{app:lemma1}
\noindent \textit{Proof.} We first note that the matrix identity,
\begin{align}
    \sum_{i=1}^{d} \Vec{m}_i \otimes \Vec{m}_i = \sum_{i=1}^{d} \Vec{n}_i \otimes \Vec{n}_i,
    \label{eq:lemma_mat_id}
\end{align}
implies the vector identity,
\begin{align}
    \sum_{i=1}^{d} \Vec{m}_i = \sum_{i=1}^{d} \Vec{n}_i,
    \label{eq:lemma_vec_id}
\end{align}
when restricted to the diagonal. In the $d=1$ case, $\Vec{m}_1 = \Vec{n}_1$ holds trivially from \cref{eq:lemma_vec_id}.

For the case $d=2$, we make the following observation from \cref{eq:lemma_vec_id}: if $\Vec{m}_1 = \Vec{n}_1$, it must follow that $\Vec{m}_2 = \Vec{n}_2$. Hence, we can assume that $\Vec{m}_1 \neq \Vec{n}_1$, and it remains only to show that $\Vec{m}_1 = \Vec{n}_2$ or $\Vec{n}_1 = \Vec{m}_2$, because in either case \cref{eq:lemma_vec_id} then immediately indicates $(\Vec{m}_1,\Vec{m}_2) = (\Vec{n}_2,\Vec{n}_1)$. By the assumption of $\Vec{m}_1 \neq \Vec{n}_1$, there exists some index $a \in \{1,\ldots,  Q \}$ such that either $(m_{1,a}, n_{1,a}) = (0, 1)$ or
$(m_{1,a}, n_{1,a}) = (1, 0)$. In these respective cases, \cref{eq:lemma_vec_id} then implies the component identity $(m_{2,a}, n_{2,a}) = (1, 0)$ or $(m_{2,a}, n_{2,a}) = (0, 1)$, respectively. By taking the $a$-th column of \cref{eq:lemma_mat_id}, we obtain 
\begin{align}
    m_{1,a} \Vec{m}_1 + m_{2,a} \Vec{m}_2 = n_{1,a} \Vec{n}_1 + n_{2,a} \Vec{n}_2.
\end{align}
Accordingly we must have either $\Vec{n}_1 = \Vec{m}_2$ or $\Vec{m}_1 = \Vec{n}_2$ as was to be shown.

Finally we turn to the case $d=3$, where we can apply our previous results inductively. First we observe that if $\Vec{m}_1 = \Vec{n}_1$, then by \cref{eq:lemma_mat_id} and our proof in the case $d=2$, it follows that $(\Vec{m}_2, \Vec{m}_3) = \mathcal{P}_2( \Vec{n}_2, \Vec{n}_3)$ for a suitable permutation $\mathcal{P}_2$.

Hence, we can assume that $\Vec{m}_1 \neq \Vec{n}_1$, so there exists an index $a \in \{ 1, \ldots , Q \}$ such that $(m_{1,a}, n_{1,a}) = (0, 1)$ or
$(m_{1,a}, n_{1,a}) = (1, 0)$. Without loss of generality, we focus on the case $(m_{1,a}, n_{1,a}) = (1, 0)$, since similar arguments hold for the complementary case $(m_{1,a}, n_{1,a}) = (0, 1)$ due to the binary symmetry.

The $a$-th component of the vector identity~\cref{eq:lemma_vec_id} then reads as 
\begin{align}
    1 + m_{2,a} + m_{3,a} = n_{2,a} + n_{3,a}.
\end{align}
Among $m_{2,a}, m_{3,a}, n_{2,a}, n_{3,a} \in \{ 0, 1 \}$, this equation is solved only be the following choices: 
\begin{align}
\label{eq:cases}
    (m_{2,a}, m_{3,a}, n_{2,a}, n_{3,a}) = \begin{cases}
        (0, 0, 1, 0) \\
        (0, 0, 0, 1) \\
        (1, 0, 1, 1) \\
        (0, 1, 1, 1)
    \end{cases}.
\end{align}

Once again taking the $a$-th column of \cref{eq:lemma_mat_id}, observe that the first two cases indicate $\Vec{m}_1 = \Vec{n}_2$ and $\Vec{m}_1 = \Vec{n}_3$, respectively. Either way, we can again reduce to the case $d=2$.

Then it remains only to consider the last two cases from \cref{eq:cases}. Taking the $a$-th column of \cref{eq:lemma_mat_id}, these two cases imply that $\Vec{m}_1 + \Vec{m}_2 = \Vec{n}_2 + \Vec{n}_3$ and $\Vec{m}_1 + \Vec{m}_3 = \Vec{n}_2 + \Vec{n}_3$, respectively. But from \cref{eq:lemma_vec_id}, we must obtain $\Vec{m}_3 = \Vec{n}_1$ and $\Vec{m}_2 = \Vec{n}_1$, respectively. Either way, we once again reduce to the case $d=2$, and the proof is complete. $\square$

We remark that these combinatorial arguments do not extend to $d \geq 4$, because the list of cases analogous to \cref{eq:cases} no longer allows a full reduction to the cases $d \leq 3$. For example under our assumption $(m_{1,a}, n_{1,a}) = (1,0)$, the relevant component-wise identity for $d=4$,
\begin{align}
    1 + m_{2,a} + m_{3,a} + m_{4,a} = n_{2,a} + n_{3,a} + n_{4,a},
\end{align}
admits many solutions, including the particular choice of $(m_{2,a}, m_{3,a}, m_{4,a}, n_{2,a}, n_{3,a}, n_{4,a}) = (1,0,0,1,1,0)$. Now evaluating the $a$-th column of \cref{eq:lemma_mat_id} with this solution, we arrive at the identities $\Vec{m}_1 + \Vec{m}_2 = \Vec{n}_2 + \Vec{n}_3$ and $\Vec{m}_3 + \Vec{m}_4 = \Vec{n}_1 + \Vec{n}_4$. However, these two vector identities do not necessarily imply the condition $(\Vec{m}_1, \Vec{m}_2) = \mathcal{P}_2 (\Vec{n}_2, \Vec{n}_3)$ or $(\Vec{m}_3, \Vec{m}_4) = \mathcal{P}_2 (\Vec{n}_1, \Vec{n}_4)$. A concrete instance to explicitly verify, even for $Q=3$, is given by  
\begin{align}
    \begin{cases}
         \Vec{m}_1 = (1,0,0) \\
         \Vec{m}_2 = (0,1,0) \\
         \Vec{m}_3 = (0,0,1) \\
         \Vec{m}_4 = (1,1,1) \\
    \end{cases}, ~~~~~\begin{cases}
         \Vec{n}_1 = (1,0,1) \\
         \Vec{n}_2 = (0,0,0) \\
         \Vec{n}_3 = (1,1,0) \\
         \Vec{n}_4 = (0,1,1)
    \end{cases},
\end{align}
where $(\Vec{m}_1, \cdots, \Vec{m}_4) \neq \mathcal{P}_4 (\Vec{n}_1, \cdots, \Vec{n}_4)$ for $\mathcal{P}_4 \in S_{4}$. This confirms that the lemma does not hold for $d=4$.

\subsection{Proof of Corollary 1}
\label{app:corollary1}
\noindent \textit{Proof.} For any $k \in \mathbb{Z}$, if $\gamma \sim \mathrm{Uniform}\{\frac{2\pi}{d+1}f: f \in \mathbb{Z}_{d+1} \}$, then the expectation
\begin{align}
    \mathbb{E} [e^{i k \gamma}] = \begin{cases}
      1, & k \in (d+1)\mathbb{Z}  \\
      0, & {\rm otherwise}
    \end{cases}
\end{align}
is nonzero only if $k$ is an integer multiple of $d+1$. Under our construction, 
\begin{align}
    k = \sum_{i=1}^{d} \bigg[\prod_{a \in \mathcal{A}} m_{i,a} - \prod_{a \in \mathcal{A}} n_{i,a} \bigg],
\end{align}
for a subset $\mathcal{A} \subseteq \{1,\cdots,Q\}$ with $\lvert \mathcal{A} \rvert = r$. Since $\lvert k \rvert \leq d$, the expectation remains non-vanishing only if $k=0$, thus replicating the statistics of $\gamma \sim \mathrm{Uniform}(0,2\pi)$. 

Let us first examine the 3-local Hamiltonian $G$ defined in \cref{eq:G_4design} with $r=3$. 
 Consequently, the statistical moments of the quantum Hutchinson state generated under this Hamiltonian can be computed as,
\begin{align}
     \mathbb{E}\left[ \left( \ket{\rchi} \bra{\rchi} \right)^{\otimes d} \right] 
     = \frac{1}{N^d} \sum_{ \{\Vec{m}_i, \Vec{n}_i \}_{i=1}^{d}}^{\rm restricted}  \ket{ \{\Vec{m}_i \} } \bra{ \{\Vec{n}_i \} }, \label{eq:1st_moment_4design} 
\end{align}
where the summation on the RHS is restricted to terms satisfying
\begin{align}
    \sum_{i=1}^{d} \Vec{m}_{i} \otimes \Vec{m}_{i} \otimes \Vec{m}_{i} = \sum_{i=1}^{d} \Vec{n}_{i} \otimes \Vec{n}_{i} \otimes \Vec{n}_{i},
    \label{eq:restriction_4design}
\end{align}
with $\otimes$ understood as a general Kronecker product such that each $\Vec{m}_{i} \otimes \Vec{m}_{i} \otimes \Vec{m}_{i}$ can be equivalently viewed as a 3-tensor with entries $(\Vec{m}_{i} \otimes \Vec{m}_{i} \otimes \Vec{m}_{i})_{abc} = m_{i,a} m_{i,b} m_{i,c}$. We notice that \cref{eq:restriction_4design} immediately implies the matrix and vector identities \cref{eq:lemma_mat_id,eq:lemma_vec_id} from Lemma 1. Our first goal is to establish that \cref{eq:restriction_4design} for $1 \leq d \leq 4$ holds precisely if and only if $(\Vec{m}_1, \cdots, \Vec{m}_d) = \mathcal{P}_d (\Vec{n}_1, \cdots, \Vec{n}_d)$ for some $\mathcal{P}_d \in S_d$.

In the case $d=1$, $\Vec{m}_1 = \Vec{n}_1$ holds trivially from exactly \cref{eq:lemma_vec_id} as in \cref{app:lemma1}.

In the cases $d=2$ and $d=3$, we assume as usual that $\Vec{m}_1 \neq \Vec{n}_1$ where $(m_{1,a}, n_{1,a}) = (1, 0)$ for some qubit index $a \in \{ 1, \ldots , Q \}$. We can then derive desired conclusions in these cases solely from the lower-rank equalities \cref{eq:lemma_mat_id,eq:lemma_vec_id}, following the proof of Theorem 1.

For the interesting case $d=4$, we once again examine the assumption $(m_{1,a}, n_{1,a}) = (1, 0)$. Let us take the $a$-th slice (column-depth slice) of the tensor identity \cref{eq:restriction_4design} by fixing the first tensor index, i.e.,
\begin{align}
     \sum_{i=1}^{4} m_{i,a} \Vec{m}_i \otimes \Vec{m}_i  = \sum_{i=1}^{4} n_{i,a} \Vec{n}_i \otimes \Vec{n}_i,
\end{align}
with possible choices,
\begin{align}
\label{eq:cases_cor}
    (m_{2,a}, m_{3,a}, m_{4,a}, n_{2,a}, n_{3,a}, n_{4,a}) = \begin{cases}
        (0, 0, 0, 1, 0, 0) \\
        (0, 0, 0, 0, 1, 0) \\
        (0, 0, 0, 0, 0, 1) \\
        (1, 0, 0, 1, 1, 0) \\
        (1, 0, 0, 1, 0, 1) \\
        (1, 0, 0, 0, 1, 1) \\
        (0, 1, 0, 1, 1, 0) \\
        (0, 1, 0, 1, 0, 1) \\
        (0, 1, 0, 0, 1, 1) \\
        (0, 0, 1, 1, 1, 0) \\
        (0, 0, 1, 1, 0, 1) \\
        (0, 0, 1, 0, 1, 1) \\
        (1, 1, 0, 1, 1, 1) \\        
        (1, 0, 1, 1, 1, 1) \\
        (0, 1, 1, 1, 1, 1) \\
    \end{cases}.
\end{align}
Now, we realize that Theorem 1 can be directly applied to all the choices above, allowing us to conclude the proof for $d=4$.

We next establish that these new quantum Hutchinson states indeed form a diagonal $7$-design, based on what we have derived thus far. For $5 \leq d \leq 7$, we want to show
\begin{align}
    \sum_{i=1}^{d} \Vec{m}_{i}^{\otimes 3} = \sum_{i=1}^{d} \Vec{n}_{i}^{\otimes 3} \iff (\Vec{m}_i)_{i=1}^{d} = \mathcal{P}_d ((\Vec{n}_i)_{i=1}^{d}),
    \label{eq:results_to_show_4designs}
\end{align}
where $\Vec{m}_{i}^{\otimes r}$ is shorthand for the $r$-fold Kronecker product of $\Vec{m}_{i}$ with itself. When $d = 5$, our starting assumption of $(m_{1,a}, n_{1,a}) = (1, 0)$ essentially yields two complementary 2-tensor identities upon restricting the first index of the 3-tensor identity,
\begin{align}
     \sum_{i \in \mathcal{I}_\alpha} \Vec{m}_{i}^{\otimes 2} = \sum_{j \in \mathcal{J}_\alpha} \Vec{n}_{j}^{\otimes 2}, \hspace{2em} \alpha = 1,2
\end{align}
where $\lvert \mathcal{I}_{\alpha} \rvert = \lvert \mathcal{J}_{\alpha} \rvert \geq 1$ by \cref{eq:lemma_vec_id}, and meanwhile $\mathcal{I}_1 \cup \mathcal{I}_2 = \mathcal{J}_1 \cup \mathcal{J}_2 = \{1,2,3,4,5\}$ form disjoint partitions. We observe that if $2 \leq \lvert \mathcal{I}_{1} \rvert \leq 3$, then Theorem 1 immediately implies $(\Vec{m}_i)_{i\in \mathcal{I}_\alpha} = \mathcal{P}_{\lvert \mathcal{J}_\alpha \rvert} ((\Vec{n}_j)_{j \in \mathcal{J}_\alpha})$. Hence, it suffices to consider the remaining cases, where we assume $\lvert \mathcal{I}_{1} \rvert = 1$ without loss of generality (otherwise we simply exchange the members $\mathcal{I}_{1} \leftrightarrow \mathcal{I}_{2}$). In this case, the 2-tensor identity over $\mathcal{I}_1$ and $\mathcal{J}_1$ trivially implies $\Vec{m}_{\mathcal{I}_{1}} = \Vec{m}_{\mathcal{J}_{1}}$, which, when combined with the LHS of \cref{eq:results_to_show_4designs}, further implies that 
\begin{align}
    \sum_{i \in \mathcal{I}_2} \Vec{m}_{i}^{\otimes 3} = \sum_{j \in \mathcal{J}_2} \Vec{n}_{i}^{\otimes 3}.
    \label{eq:3tensor_5designs}
\end{align}
However, we can now use our result for $d=4$ since $\lvert \mathcal{I}_2 \rvert = \lvert \mathcal{J}_2 \rvert = 4$, which allows us to conclude for $d=5$ (thus now \cref{eq:results_to_show_4designs} is valid when $1 \leq d \leq 5$). For $d=6$, we almost repeat the same analysis, except that we need to consider the cases where $\lvert \mathcal{I}_{1} \rvert =1$ and $\lvert \mathcal{I}_{1} \rvert =2$. If $\lvert \mathcal{I}_{1} \rvert =1$, then we can directly invoke the updated $d=5$ result to conclude. If $\lvert \mathcal{I}_{1} \rvert =2$, the relevant 2-tensor identity over $\mathcal{I}_1$ and $\mathcal{J}_1$ leads to a 3-tensor identity of the form \cref{eq:3tensor_5designs} over $\mathcal{I}_2$ and $\mathcal{J}_2$ with $\lvert \mathcal{I}_2 \rvert = \lvert \mathcal{J}_2 \rvert = 4$, allowing us to fully conclude for $d=6$. Finally for $d=7$, we encounter the additional nontrivial case of $\lvert \mathcal{I}_1 \rvert = 3$ and $\lvert \mathcal{I}_2 \rvert = 4$. Once again, the relevant 2-tensor identity over $\mathcal{I}_1$ and $\mathcal{J}_1$ helps us deduce a 3-tensor identity of the form \cref{eq:3tensor_5designs} over $\mathcal{I}_2$ and $\mathcal{J}_2$ with $\lvert \mathcal{I}_2 \rvert = \lvert \mathcal{J}_2 \rvert = 4$, which therefore concludes the proof for $4 \leq d \leq 7$. At this point, we can already see why such argument breaks down for $d=8$: the case $\lvert \mathcal{I}_{1} \rvert = \lvert \mathcal{I}_{2} \rvert = 4$ exhausts the utility of our 2-tensor (low-rank) identities, which only provide us information if $\lvert \mathcal{I}_{1} \rvert \leq 3$ or $\lvert \mathcal{I}_{2} \rvert \leq 3$.

This completes the argument for $r=2$ and motivates the introduction of higher-rank tensor identities for achieving higher design orders $d$ by considering higher interaction locality $r>3$.

In general, for $r>3$, we want to establish the identity
\begin{align}
    \sum_{i=1}^{d} \Vec{m}_{i}^{\otimes r} = \sum_{i=1}^{d} \Vec{n}_{i}^{\otimes r} \iff (\Vec{m}_i)_{i=1}^{d} = \mathcal{P}_d ((\Vec{n}_i)_{i=1}^{d}),
\label{eq:results_to_show_generaldesigns}
\end{align}
for all $d=1,\ldots, 2^r - 1$.

Assume that the equation on the LHS of \eqref{eq:results_to_show_generaldesigns} holds,
i.e., that
\begin{equation}
    \sum_{i=1}^{d} \Vec{m}_{i}^{\otimes r} = \sum_{i=1}^{d} \Vec{n}_{i}^{\otimes r}.
    \label{eq:blah}
\end{equation}
Then in particular,
\begin{align}
    \sum_{i=1}^{d} \Vec{m}_{i}^{\otimes (r-1)} = \sum_{i=1}^{d} \Vec{n}_{i}^{\otimes (r-1)}.
    \label{eq:rminus1results_generaldesigns}
\end{align}
By induction on $r$, for any $d=1,\ldots, 2^{r-1} - 1$, it follows that 
\begin{align}
    (\Vec{m}_i)_{i=1}^{d} = \mathcal{P}_d ((\Vec{n}_i)_{i=1}^{d}).
\end{align}
Therefore we can assume without loss of generality that $d \geq 2^{r-1}$.

Let us then extend the validity of \cref{eq:results_to_show_generaldesigns} for increasingly large $d$, starting with the case $d=2^{r-1}$. We may assume without loss of generality that $\Vec{m}_i \neq \Vec{n}_i$ for some $i \in \{1, \ldots, d\}$. (Otherwise, we can reduce to the case of $d < 2^{r-1}$.) Then there exists a qubit index $a \in \{ 1, \ldots , Q \}$ such that $m_{i,a} \neq n_{i,a}$. Upon reordering the $i$ indices and exchanging the roles of $\Vec{m}$ and $\Vec{n}$, we may simply assume that $m_{1,a} = 1$ and $n_{1,a} = 0$.

By considering the $(a,\ldots,a)$-th entry of the equation \eqref{eq:blah} of $r$-index tensors, we deduce that 
\begin{align}
    \sum_{i=1}^{d} {m}_{i,a} = \sum_{i=1}^{d} {n}_{i,a}.
    \label{eq:scalar_sum_to_show_generaldesigns}
\end{align}
Since $(m_{1,a}, n_{1,a}) = (1,0)$, \cref{eq:scalar_sum_to_show_generaldesigns} implies that there must exist some $j \neq 1$ such that $(m_{j,a}, n_{j,a}) = (0,1)$. Let
\begin{subequations}
\begin{align}
    \mathcal{I}_{1} &= \{ i: m_{i,a} = 1 \}, \label{eq:I1_generaldesigns} \\
    \mathcal{J}_{1} &= \{ i: n_{i,a} = 1 \}.
    \label{eq:J1_generaldesigns}
\end{align}
\end{subequations}
We note that $1 \leq \lvert \mathcal{I}_{1} \rvert = \lvert \mathcal{J}_{1} \rvert \leq d-1 = 2^{r-1}-1$ (although the sets $\mathcal{I}_{1}$ and $\mathcal{J}_{1}$ need not be identical).

\cref{eq:blah} is an equation of two $r$-index tensors. Now We consider the equation of $(r-1)$-index tensors induced by setting the first index of both sides to be $a$:
\begin{align}
    \sum_{i \in \mathcal{I}_1 }^{d} \Vec{m}_{i}^{\otimes (r-1)} = \sum_{ i \in \mathcal{J}_1}^{d} \Vec{n}_{i}^{\otimes (r-1)}.
    \label{eq:rminus1_sum_to_show_generaldesigns}
\end{align}
% where the number of nonzero terms in each sum is exactly $\lvert \mathcal{I}_{1} \rvert = \lvert \mathcal{J}_{1} \rvert \leq 2^{r-1} - 1$.
Since both sides are sums over at most $2^{r-1} - 1$ terms, we deduce by induction that 
\begin{align}
    (\Vec{m}_i)_{i \in \mathcal{I}_1} = \mathcal{P}_{\lvert \mathcal{J}_1 \rvert} ((\Vec{n}_i)_{i \in \mathcal{J}_1}).
    \label{eq:sum_I1_show_generaldesigns}
\end{align}
(Henceforth we will adopt the shorthand $(\Vec{m}_i)_{\mathcal{I}_1}$ for, \textit{e.g.}, the LHS of the last equation.)

Moreover, \cref{eq:sum_I1_show_generaldesigns} immediately implies that 
\begin{align}
    \sum_{i \in \mathcal{I}_1} \Vec{m}_{i}^{\otimes r} = \sum_{i \in \mathcal{J}_1} \Vec{n}_{i}^{\otimes r}.
    \label{eq:r_I1J1_to_show_generaldesigns}
\end{align}
Combining this with \cref{eq:blah}, we obtain
\begin{align}
    \sum_{i \in \mathcal{I}_2} \Vec{m}_{i}^{\otimes r} = \sum_{i \in \mathcal{J}_2} \Vec{n}_{i}^{\otimes r},
    \label{eq:r_I2J2_to_show_generaldesigns}
\end{align}
where
\begin{subequations}
\begin{align}
    \mathcal{I}_{2} &= \{ i: m_{i,a} = 0 \}, \\
    \mathcal{J}_{2} &= \{ i: n_{i,a} = 0 \},
\end{align}
\end{subequations}
with $1 \leq \lvert \mathcal{I}_{2} \lvert  = \lvert \mathcal{J}_{2} \lvert \leq 2^{r-1} - 1$ (since $\lvert \mathcal{I}_{1} \rvert + \lvert \mathcal{I}_{2} \rvert = \lvert \mathcal{J}_{1} \rvert +  \lvert \mathcal{J}_{2} \rvert = d$). Thus applying induction once again, we arrive at the decomposition, 
\begin{align}
    ((\Vec{m}_i)_{\mathcal{I}_1}, (\Vec{m}_i)_{\mathcal{I}_2} ) = (\mathcal{P}_{\lvert \mathcal{J}_1 \rvert} ((\Vec{n}_i)_{ \mathcal{J}_1}), \mathcal{P}_{\lvert \mathcal{J}_2 \rvert} ((\Vec{n}_i)_{\mathcal{J}_2})),
    \label{eq:decomp}
\end{align}
which verifies \cref{eq:results_to_show_generaldesigns} for the case $d = 2^{r-1}$.

Next, we consider the case $d = 2^{r-1}+1$. As before, we assume $m_{1,a} = 1$ and $n_{1,a} = 0$ for some $a \in \{1,\ldots,Q\}$; otherwise we can reduce to the previous case of $d = 2^{r-1}$. Following the same reasoning through \cref{eq:scalar_sum_to_show_generaldesigns,eq:I1_generaldesigns,eq:J1_generaldesigns}, we likewise obtain \cref{eq:rminus1_sum_to_show_generaldesigns}, this time with $1 \leq \lvert \mathcal{I}_{1} \rvert \leq d-1 = 2^{r-1}$. 

If $\lvert \mathcal{I}_{1} \rvert < 2^{r-1}$, again we recover $ (\Vec{m}_i)_{\mathcal{I}_1} = \mathcal{P}_{\lvert \mathcal{J}_1 \rvert} ((\Vec{n}_i)_{\mathcal{J}_1})$ just as in \cref{eq:sum_I1_show_generaldesigns}. This implies \cref{eq:r_I1J1_to_show_generaldesigns}, and thereby \cref{eq:r_I2J2_to_show_generaldesigns} with $\lvert \mathcal{I}_{2} \rvert \leq d-1 = 2^{r-1}$. Then by induction on the sets $\mathcal{I}_2$ and $\mathcal{J}_2$, we can deduce $ (\Vec{m}_i)_{\mathcal{I}_2} = \mathcal{P}_{\lvert \mathcal{J}_2 \rvert} ((\Vec{n}_i)_{\mathcal{J}_2})$ and accordingly \cref{eq:results_to_show_generaldesigns} for $d = 2^{r-1}+1$.

If instead $\lvert \mathcal{I}_{1} \rvert = 2^{r-1}$, we have $\lvert \mathcal{I}_2 \rvert = d - \lvert \mathcal{I}_1 \rvert = 1$ and additionally, by \cref{eq:rminus1_sum_to_show_generaldesigns,eq:rminus1results_generaldesigns},
\begin{align}
    \sum_{i \in \mathcal{I}_2} \Vec{m}_{i}^{\otimes (r-1)} = \sum_{i \in \mathcal{J}_2} \Vec{n}_{i}^{\otimes (r-1)}. 
\end{align}
This identity implies via induction that $ (\Vec{m}_i)_{\mathcal{I}_2} = \mathcal{P}_{\lvert \mathcal{J}_2 \rvert} ((\Vec{n}_i)_{\mathcal{J}_2})$ and hence \cref{eq:r_I2J2_to_show_generaldesigns}, which further implies \cref{eq:r_I1J1_to_show_generaldesigns}. By induction on the size of the sets $\mathcal{I}_1$ and $\mathcal{J}_1$, we then deduce $(\Vec{m}_i)_{\mathcal{I}_1} = \mathcal{P}_{\lvert \mathcal{J}_1 \rvert} ((\Vec{n}_i)_{\mathcal{J}_1})$, so that \cref{eq:decomp} again holds. This verifies \cref{eq:results_to_show_generaldesigns} for $d = 2^{r-1}+1$.

We can continue increasing $d$ and employing the same proof strategy (essentially \cref{eq:rminus1results_generaldesigns,eq:rminus1_sum_to_show_generaldesigns}, and our inductive hypothesis) iteratively, until we encounter the possibility that both $\lvert \mathcal{I}_1 \rvert \geq 2^{r-1}$ and $\lvert \mathcal{I}_2 \rvert \geq 2^{r-1}$. That is, we proceed unless the sizes of these subsets exceed the threshold where the supporting $(r-1)$-tensor identity no longer applies directly. Concretely, this allows us to reach $d = 2^r - 1$, as desired.  $\square$

% \rev{In general, an $r$-local generating Hamiltonian $G$ gives us access to an established relation of \cref{eq:results_to_show_4designs} with an $(r-1)$-fold Kronecker product, while also allowing us to promote it to the desired relation for higher $d$ involving an $r$-fold Kronecker product.} 

% Suppose the $(r-1)$-tensor identity leads to the desired combinatorial identity $(\Vec{m}_i)_{i=1}^{d} = \mathcal{P}_d ((\Vec{n}_i)_{i=1}^{d})$, up to some order $d \leq d_{r-1}$. Then its power is exhausted when $\lvert \mathcal{I}_1 \rvert \geq d_{r-1} + 1$ and $\lvert \mathcal{I}_2 \rvert \geq d_{r-1}+1$, which occurs if $d \geq 2d_{r-1}+2$. This gives us a recursive recipe for finding $d_r$ with
% \begin{align}
%     d_r = \begin{cases} 1 & r=1 \\
%                         3 & r=2 \\
%                         7 & r=3
%         \end{cases},
% \end{align}
% and therefore $d_r = 2^{r}-1$ for $r \geq 4$.

% In turn, our construction of higher-order diagonal $d$-designs proceeds by systematically increasing the locality of interactions in the generating Hamiltonian $G$, which naturally produces higher-rank tensor identities that can be analyzed with lower-rank ones recursively. $\square$

% \bibliographystyle{apsrev4-1}
\bibliography{mainbib.bib}

%merlin.mbs apsrev4-1.bst 2010-07-25 4.21a (PWD, AO, DPC) hacked
%Control: key (0)
%Control: author (0) dotless jnrlst
%Control: editor formatted (1) identically to author
%Control: production of article title (0) allowed
%Control: page (1) range
%Control: year (0) verbatim
%Control: production of eprint (0) enabled
\begin{thebibliography}{38}%
\makeatletter
\providecommand \@ifxundefined [1]{%
 \@ifx{#1\undefined}
}%
\providecommand \@ifnum [1]{%
 \ifnum #1\expandafter \@firstoftwo
 \else \expandafter \@secondoftwo
 \fi
}%
\providecommand \@ifx [1]{%
 \ifx #1\expandafter \@firstoftwo
 \else \expandafter \@secondoftwo
 \fi
}%
\providecommand \natexlab [1]{#1}%
\providecommand \enquote  [1]{``#1''}%
\providecommand \bibnamefont  [1]{#1}%
\providecommand \bibfnamefont [1]{#1}%
\providecommand \citenamefont [1]{#1}%
\providecommand \href@noop [0]{\@secondoftwo}%
\providecommand \href [0]{\begingroup \@sanitize@url \@href}%
\providecommand \@href[1]{\@@startlink{#1}\@@href}%
\providecommand \@@href[1]{\endgroup#1\@@endlink}%
\providecommand \@sanitize@url [0]{\catcode `\\12\catcode `\$12\catcode `\&12\catcode `\#12\catcode `\^12\catcode `\_12\catcode `\%12\relax}%
\providecommand \@@startlink[1]{}%
\providecommand \@@endlink[0]{}%
\providecommand \url  [0]{\begingroup\@sanitize@url \@url }%
\providecommand \@url [1]{\endgroup\@href {#1}{\urlprefix }}%
\providecommand \urlprefix  [0]{URL }%
\providecommand \Eprint [0]{\href }%
\providecommand \doibase [0]{http://dx.doi.org/}%
\providecommand \selectlanguage [0]{\@gobble}%
\providecommand \bibinfo  [0]{\@secondoftwo}%
\providecommand \bibfield  [0]{\@secondoftwo}%
\providecommand \translation [1]{[#1]}%
\providecommand \BibitemOpen [0]{}%
\providecommand \bibitemStop [0]{}%
\providecommand \bibitemNoStop [0]{.\EOS\space}%
\providecommand \EOS [0]{\spacefactor3000\relax}%
\providecommand \BibitemShut  [1]{\csname bibitem#1\endcsname}%
\let\auto@bib@innerbib\@empty
%</preamble>
\bibitem [{\citenamefont {Hayden}\ \emph {et~al.}(2004)\citenamefont {Hayden}, \citenamefont {Leung}, \citenamefont {Shor},\ and\ \citenamefont {Winter}}]{Hayden2004}%
  \BibitemOpen
  \bibfield  {author} {\bibinfo {author} {\bibfnamefont {Patrick}\ \bibnamefont {Hayden}}, \bibinfo {author} {\bibfnamefont {Debbie}\ \bibnamefont {Leung}}, \bibinfo {author} {\bibfnamefont {Peter~W.}\ \bibnamefont {Shor}}, \ and\ \bibinfo {author} {\bibfnamefont {Andreas}\ \bibnamefont {Winter}},\ }\bibfield  {title} {\enquote {\bibinfo {title} {Randomizing quantum states: Constructions and applications},}\ }\href {\doibase 10.1007/s00220-004-1087-6} {\bibfield  {journal} {\bibinfo  {journal} {Communications in Mathematical Physics}\ }\textbf {\bibinfo {volume} {250}},\ \bibinfo {pages} {371--391} (\bibinfo {year} {2004})}\BibitemShut {NoStop}%
\bibitem [{\citenamefont {Dankert}\ \emph {et~al.}(2009)\citenamefont {Dankert}, \citenamefont {Cleve}, \citenamefont {Emerson},\ and\ \citenamefont {Livine}}]{Fidelity}%
  \BibitemOpen
  \bibfield  {author} {\bibinfo {author} {\bibfnamefont {Christoph}\ \bibnamefont {Dankert}}, \bibinfo {author} {\bibfnamefont {Richard}\ \bibnamefont {Cleve}}, \bibinfo {author} {\bibfnamefont {Joseph}\ \bibnamefont {Emerson}}, \ and\ \bibinfo {author} {\bibfnamefont {Etera}\ \bibnamefont {Livine}},\ }\bibfield  {title} {\enquote {\bibinfo {title} {Exact and approximate unitary 2-designs and their application to fidelity estimation},}\ }\href {\doibase 10.1103/PhysRevA.80.012304} {\bibfield  {journal} {\bibinfo  {journal} {Phys. Rev. A}\ }\textbf {\bibinfo {volume} {80}},\ \bibinfo {pages} {012304} (\bibinfo {year} {2009})}\BibitemShut {NoStop}%
\bibitem [{\citenamefont {Huang}\ \emph {et~al.}(2020)\citenamefont {Huang}, \citenamefont {Kueng},\ and\ \citenamefont {Preskill}}]{Preskill_shadow}%
  \BibitemOpen
  \bibfield  {author} {\bibinfo {author} {\bibfnamefont {Hsin-Yuan}\ \bibnamefont {Huang}}, \bibinfo {author} {\bibfnamefont {Richard}\ \bibnamefont {Kueng}}, \ and\ \bibinfo {author} {\bibfnamefont {John}\ \bibnamefont {Preskill}},\ }\bibfield  {title} {\enquote {\bibinfo {title} {Predicting many properties of a quantum system from very few measurements},}\ }\href {\doibase 10.1038/s41567-020-0932-7} {\bibfield  {journal} {\bibinfo  {journal} {Nature Physics}\ }\textbf {\bibinfo {volume} {16}},\ \bibinfo {pages} {1050--1057} (\bibinfo {year} {2020})}\BibitemShut {NoStop}%
\bibitem [{\citenamefont {Mele}(2024)}]{Mele2024introductiontohaar}%
  \BibitemOpen
  \bibfield  {author} {\bibinfo {author} {\bibfnamefont {Antonio~Anna}\ \bibnamefont {Mele}},\ }\bibfield  {title} {\enquote {\bibinfo {title} {Introduction to {H}aar {M}easure {T}ools in {Q}uantum {I}nformation: {A} {B}eginner's {T}utorial},}\ }\href {\doibase 10.22331/q-2024-05-08-1340} {\bibfield  {journal} {\bibinfo  {journal} {{Quantum}}\ }\textbf {\bibinfo {volume} {8}},\ \bibinfo {pages} {1340} (\bibinfo {year} {2024})}\BibitemShut {NoStop}%
\bibitem [{\citenamefont {Nakata}\ and\ \citenamefont {Murao}(2013)}]{diagonal_2013}%
  \BibitemOpen
  \bibfield  {author} {\bibinfo {author} {\bibfnamefont {Yoshifumi}\ \bibnamefont {Nakata}}\ and\ \bibinfo {author} {\bibfnamefont {Mio}\ \bibnamefont {Murao}},\ }\bibfield  {title} {\enquote {\bibinfo {title} {Diagonal-unitary 2-design and their implementations by quantum circuits},}\ }\href {\doibase 10.1142/S0219749913500627} {\bibfield  {journal} {\bibinfo  {journal} {International Journal of Quantum Information}\ }\textbf {\bibinfo {volume} {11}},\ \bibinfo {pages} {1350062} (\bibinfo {year} {2013})}\BibitemShut {NoStop}%
\bibitem [{\citenamefont {Nakata}\ \emph {et~al.}(2014)\citenamefont {Nakata}, \citenamefont {Koashi},\ and\ \citenamefont {Murao}}]{diagonal_2014}%
  \BibitemOpen
  \bibfield  {author} {\bibinfo {author} {\bibfnamefont {Yoshifumi}\ \bibnamefont {Nakata}}, \bibinfo {author} {\bibfnamefont {Masato}\ \bibnamefont {Koashi}}, \ and\ \bibinfo {author} {\bibfnamefont {Mio}\ \bibnamefont {Murao}},\ }\bibfield  {title} {\enquote {\bibinfo {title} {Generating a state t-design by diagonal quantum circuits},}\ }\href {\doibase 10.1088/1367-2630/16/5/053043} {\bibfield  {journal} {\bibinfo  {journal} {New Journal of Physics}\ }\textbf {\bibinfo {volume} {16}},\ \bibinfo {pages} {053043} (\bibinfo {year} {2014})}\BibitemShut {NoStop}%
\bibitem [{\citenamefont {Aliferis}\ \emph {et~al.}(2009)\citenamefont {Aliferis}, \citenamefont {Brito}, \citenamefont {DiVincenzo}, \citenamefont {Preskill}, \citenamefont {Steffen},\ and\ \citenamefont {Terhal}}]{Aliferis_2009}%
  \BibitemOpen
  \bibfield  {author} {\bibinfo {author} {\bibfnamefont {P}~\bibnamefont {Aliferis}}, \bibinfo {author} {\bibfnamefont {F}~\bibnamefont {Brito}}, \bibinfo {author} {\bibfnamefont {D~P}\ \bibnamefont {DiVincenzo}}, \bibinfo {author} {\bibfnamefont {J}~\bibnamefont {Preskill}}, \bibinfo {author} {\bibfnamefont {M}~\bibnamefont {Steffen}}, \ and\ \bibinfo {author} {\bibfnamefont {B~M}\ \bibnamefont {Terhal}},\ }\bibfield  {title} {\enquote {\bibinfo {title} {Fault-tolerant computing with biased-noise superconducting qubits: a case study},}\ }\href {\doibase 10.1088/1367-2630/11/1/013061} {\bibfield  {journal} {\bibinfo  {journal} {New Journal of Physics}\ }\textbf {\bibinfo {volume} {11}},\ \bibinfo {pages} {013061} (\bibinfo {year} {2009})}\BibitemShut {NoStop}%
\bibitem [{\citenamefont {Hu}\ and\ \citenamefont {You}(2022)}]{Hu2022}%
  \BibitemOpen
  \bibfield  {author} {\bibinfo {author} {\bibfnamefont {Hong-Ye}\ \bibnamefont {Hu}}\ and\ \bibinfo {author} {\bibfnamefont {Yi-Zhuang}\ \bibnamefont {You}},\ }\bibfield  {title} {\enquote {\bibinfo {title} {Hamiltonian-driven shadow tomography of quantum states},}\ }\href {\doibase 10.1103/PhysRevResearch.4.013054} {\bibfield  {journal} {\bibinfo  {journal} {Phys. Rev. Res.}\ }\textbf {\bibinfo {volume} {4}},\ \bibinfo {pages} {013054} (\bibinfo {year} {2022})}\BibitemShut {NoStop}%
\bibitem [{\citenamefont {Tran}\ \emph {et~al.}(2023)\citenamefont {Tran}, \citenamefont {Mark}, \citenamefont {Ho},\ and\ \citenamefont {Choi}}]{Tran2023}%
  \BibitemOpen
  \bibfield  {author} {\bibinfo {author} {\bibfnamefont {Minh~C.}\ \bibnamefont {Tran}}, \bibinfo {author} {\bibfnamefont {Daniel~K.}\ \bibnamefont {Mark}}, \bibinfo {author} {\bibfnamefont {Wen~Wei}\ \bibnamefont {Ho}}, \ and\ \bibinfo {author} {\bibfnamefont {Soonwon}\ \bibnamefont {Choi}},\ }\bibfield  {title} {\enquote {\bibinfo {title} {Measuring arbitrary physical properties in analog quantum simulation},}\ }\href {\doibase 10.1103/PhysRevX.13.011049} {\bibfield  {journal} {\bibinfo  {journal} {Phys. Rev. X}\ }\textbf {\bibinfo {volume} {13}},\ \bibinfo {pages} {011049} (\bibinfo {year} {2023})}\BibitemShut {NoStop}%
\bibitem [{\citenamefont {McGinley}\ and\ \citenamefont {Fava}(2023)}]{McGinley2023}%
  \BibitemOpen
  \bibfield  {author} {\bibinfo {author} {\bibfnamefont {Max}\ \bibnamefont {McGinley}}\ and\ \bibinfo {author} {\bibfnamefont {Michele}\ \bibnamefont {Fava}},\ }\bibfield  {title} {\enquote {\bibinfo {title} {Shadow tomography from emergent state designs in analog quantum simulators},}\ }\href {\doibase 10.1103/PhysRevLett.131.160601} {\bibfield  {journal} {\bibinfo  {journal} {Phys. Rev. Lett.}\ }\textbf {\bibinfo {volume} {131}},\ \bibinfo {pages} {160601} (\bibinfo {year} {2023})}\BibitemShut {NoStop}%
\bibitem [{\citenamefont {Liu}\ \emph {et~al.}(2024)\citenamefont {Liu}, \citenamefont {Hao},\ and\ \citenamefont {Hu}}]{liu2024predicting}%
  \BibitemOpen
  \bibfield  {author} {\bibinfo {author} {\bibfnamefont {Zhenhuan}\ \bibnamefont {Liu}}, \bibinfo {author} {\bibfnamefont {Zihan}\ \bibnamefont {Hao}}, \ and\ \bibinfo {author} {\bibfnamefont {Hong-Ye}\ \bibnamefont {Hu}},\ }\bibfield  {title} {\enquote {\bibinfo {title} {Predicting arbitrary state properties from single hamiltonian quench dynamics},}\ }\href {\doibase https://doi.org/10.48550/arXiv.2311.00695} {\  (\bibinfo {year} {2024}),\ https://doi.org/10.48550/arXiv.2311.00695}\BibitemShut {NoStop}%
\bibitem [{\citenamefont {Cowtan}\ \emph {et~al.}(2019)\citenamefont {Cowtan}, \citenamefont {Dilkes}, \citenamefont {Duncan}, \citenamefont {Simmons},\ and\ \citenamefont {Sivarajah}}]{phasegadget_2019}%
  \BibitemOpen
  \bibfield  {author} {\bibinfo {author} {\bibfnamefont {Alexander}\ \bibnamefont {Cowtan}}, \bibinfo {author} {\bibfnamefont {Silas}\ \bibnamefont {Dilkes}}, \bibinfo {author} {\bibfnamefont {Ross}\ \bibnamefont {Duncan}}, \bibinfo {author} {\bibfnamefont {Will}\ \bibnamefont {Simmons}}, \ and\ \bibinfo {author} {\bibfnamefont {Seyon}\ \bibnamefont {Sivarajah}},\ }\bibfield  {title} {\enquote {\bibinfo {title} {Phase gadget synthesis for shallow circuits},}\ }in\ \href {\doibase 10.4204/EPTCS.318.13} {\emph {\bibinfo {booktitle} {Proceedings of the 16th International Conference on Quantum Physics and Logic (QPL 2019)}}},\ \bibinfo {series} {Electronic Proceedings in Theoretical Computer Science}, Vol.\ \bibinfo {volume} {318},\ \bibinfo {editor} {edited by\ \bibinfo {editor} {\bibfnamefont {Bob}\ \bibnamefont {Coecke}}\ and\ \bibinfo {editor} {\bibfnamefont {Matthew}\ \bibnamefont {Leifer}}}\ (\bibinfo  {publisher} {Open Publishing Association},\ \bibinfo {year} {2019})\ pp.\ \bibinfo {pages}
  {213--228}\BibitemShut {NoStop}%
\bibitem [{\citenamefont {Litinski}(2019)}]{Litinski2019gameofsurfacecodes}%
  \BibitemOpen
  \bibfield  {author} {\bibinfo {author} {\bibfnamefont {Daniel}\ \bibnamefont {Litinski}},\ }\bibfield  {title} {\enquote {\bibinfo {title} {A {G}ame of {S}urface {C}odes: {L}arge-{S}cale {Q}uantum {C}omputing with {L}attice {S}urgery},}\ }\href {\doibase 10.22331/q-2019-03-05-128} {\bibfield  {journal} {\bibinfo  {journal} {{Quantum}}\ }\textbf {\bibinfo {volume} {3}},\ \bibinfo {pages} {128} (\bibinfo {year} {2019})}\BibitemShut {NoStop}%
\bibitem [{\citenamefont {Moflic}\ and\ \citenamefont {Paler}(2024)}]{moflic2024constantdepthimplementationpauli}%
  \BibitemOpen
  \bibfield  {author} {\bibinfo {author} {\bibfnamefont {Ioana}\ \bibnamefont {Moflic}}\ and\ \bibinfo {author} {\bibfnamefont {Alexandru}\ \bibnamefont {Paler}},\ }\href {\doibase https://doi.org/10.48550/arXiv.2408.08265} {\enquote {\bibinfo {title} {On the constant depth implementation of pauli exponentials},}\ } (\bibinfo {year} {2024})\BibitemShut {NoStop}%
\bibitem [{\citenamefont {Gross}\ \emph {et~al.}(2007)\citenamefont {Gross}, \citenamefont {Audenaert},\ and\ \citenamefont {Eisert}}]{Gross2007}%
  \BibitemOpen
  \bibfield  {author} {\bibinfo {author} {\bibfnamefont {D.}~\bibnamefont {Gross}}, \bibinfo {author} {\bibfnamefont {K.}~\bibnamefont {Audenaert}}, \ and\ \bibinfo {author} {\bibfnamefont {J.}~\bibnamefont {Eisert}},\ }\bibfield  {title} {\enquote {\bibinfo {title} {{Evenly distributed unitaries: On the structure of unitary designs}},}\ }\href {\doibase 10.1063/1.2716992} {\bibfield  {journal} {\bibinfo  {journal} {Journal of Mathematical Physics}\ }\textbf {\bibinfo {volume} {48}},\ \bibinfo {pages} {052104} (\bibinfo {year} {2007})}\BibitemShut {NoStop}%
\bibitem [{\citenamefont {Harrow}\ \emph {et~al.}(2009)\citenamefont {Harrow}, \citenamefont {Hassidim},\ and\ \citenamefont {Lloyd}}]{Harrow2009}%
  \BibitemOpen
  \bibfield  {author} {\bibinfo {author} {\bibfnamefont {Aram~W.}\ \bibnamefont {Harrow}}, \bibinfo {author} {\bibfnamefont {Avinatan}\ \bibnamefont {Hassidim}}, \ and\ \bibinfo {author} {\bibfnamefont {Seth}\ \bibnamefont {Lloyd}},\ }\bibfield  {title} {\enquote {\bibinfo {title} {Quantum algorithm for linear systems of equations},}\ }\href {\doibase 10.1103/PhysRevLett.103.150502} {\bibfield  {journal} {\bibinfo  {journal} {Phys. Rev. Lett.}\ }\textbf {\bibinfo {volume} {103}},\ \bibinfo {pages} {150502} (\bibinfo {year} {2009})}\BibitemShut {NoStop}%
\bibitem [{\citenamefont {Girard}(1989)}]{girard1989fast}%
  \BibitemOpen
  \bibfield  {author} {\bibinfo {author} {\bibfnamefont {A}~\bibnamefont {Girard}},\ }\bibfield  {title} {\enquote {\bibinfo {title} {A fast ‘monte-carlo cross-validation’procedure for large least squares problems with noisy data},}\ }\href@noop {} {\bibfield  {journal} {\bibinfo  {journal} {Numerische Mathematik}\ }\textbf {\bibinfo {volume} {56}},\ \bibinfo {pages} {1--23} (\bibinfo {year} {1989})}\BibitemShut {NoStop}%
\bibitem [{\citenamefont {Hutchinson}(1990)}]{Hutch}%
  \BibitemOpen
  \bibfield  {author} {\bibinfo {author} {\bibfnamefont {M.F.}\ \bibnamefont {Hutchinson}},\ }\bibfield  {title} {\enquote {\bibinfo {title} {A stochastic estimator of the trace of the influence matrix for laplacian smoothing splines},}\ }\href {\doibase 10.1080/03610919008812866} {\bibfield  {journal} {\bibinfo  {journal} {Communications in Statistics - Simulation and Computation}\ }\textbf {\bibinfo {volume} {19}},\ \bibinfo {pages} {433--450} (\bibinfo {year} {1990})}\BibitemShut {NoStop}%
\bibitem [{\citenamefont {Iitaka}\ and\ \citenamefont {Ebisuzaki}(2004)}]{random_phase}%
  \BibitemOpen
  \bibfield  {author} {\bibinfo {author} {\bibfnamefont {Toshiaki}\ \bibnamefont {Iitaka}}\ and\ \bibinfo {author} {\bibfnamefont {Toshikazu}\ \bibnamefont {Ebisuzaki}},\ }\bibfield  {title} {\enquote {\bibinfo {title} {Random phase vector for calculating the trace of a large matrix},}\ }\href {\doibase 10.1103/PhysRevE.69.057701} {\bibfield  {journal} {\bibinfo  {journal} {Phys. Rev. E}\ }\textbf {\bibinfo {volume} {69}},\ \bibinfo {pages} {057701} (\bibinfo {year} {2004})}\BibitemShut {NoStop}%
\bibitem [{\citenamefont {Knill}\ and\ \citenamefont {Laflamme}(1998)}]{one_clean_qubit}%
  \BibitemOpen
  \bibfield  {author} {\bibinfo {author} {\bibfnamefont {E.}~\bibnamefont {Knill}}\ and\ \bibinfo {author} {\bibfnamefont {R.}~\bibnamefont {Laflamme}},\ }\bibfield  {title} {\enquote {\bibinfo {title} {Power of one bit of quantum information},}\ }\href {\doibase 10.1103/PhysRevLett.81.5672} {\bibfield  {journal} {\bibinfo  {journal} {Phys. Rev. Lett.}\ }\textbf {\bibinfo {volume} {81}},\ \bibinfo {pages} {5672--5675} (\bibinfo {year} {1998})}\BibitemShut {NoStop}%
\bibitem [{\citenamefont {Cleve}\ \emph {et~al.}(2016)\citenamefont {Cleve}, \citenamefont {Leung}, \citenamefont {Liu},\ and\ \citenamefont {Wang}}]{Cleve2016}%
  \BibitemOpen
  \bibfield  {author} {\bibinfo {author} {\bibfnamefont {Richard}\ \bibnamefont {Cleve}}, \bibinfo {author} {\bibfnamefont {Debbie}\ \bibnamefont {Leung}}, \bibinfo {author} {\bibfnamefont {Li}~\bibnamefont {Liu}}, \ and\ \bibinfo {author} {\bibfnamefont {Chunhao}\ \bibnamefont {Wang}},\ }\bibfield  {title} {\enquote {\bibinfo {title} {Near-linear constructions of exact unitary 2-designs},}\ }\href@noop {} {\bibfield  {journal} {\bibinfo  {journal} {Quantum Info. Comput.}\ }\textbf {\bibinfo {volume} {16}},\ \bibinfo {pages} {721–756} (\bibinfo {year} {2016})}\BibitemShut {NoStop}%
\bibitem [{\citenamefont {Nakata}\ \emph {et~al.}(2017)\citenamefont {Nakata}, \citenamefont {Hirche}, \citenamefont {Koashi},\ and\ \citenamefont {Winter}}]{NHKW17}%
  \BibitemOpen
  \bibfield  {author} {\bibinfo {author} {\bibfnamefont {Yoshifumi}\ \bibnamefont {Nakata}}, \bibinfo {author} {\bibfnamefont {Christoph}\ \bibnamefont {Hirche}}, \bibinfo {author} {\bibfnamefont {Masato}\ \bibnamefont {Koashi}}, \ and\ \bibinfo {author} {\bibfnamefont {Andreas}\ \bibnamefont {Winter}},\ }\bibfield  {title} {\enquote {\bibinfo {title} {Efficient quantum pseudorandomness with nearly time-independent hamiltonian dynamics},}\ }\href {\doibase 10.1103/PhysRevX.7.021006} {\bibfield  {journal} {\bibinfo  {journal} {Phys. Rev. X}\ }\textbf {\bibinfo {volume} {7}},\ \bibinfo {pages} {021006} (\bibinfo {year} {2017})}\BibitemShut {NoStop}%
\bibitem [{\citenamefont {Haah}\ \emph {et~al.}(2024)\citenamefont {Haah}, \citenamefont {Liu},\ and\ \citenamefont {Tan}}]{HLT24}%
  \BibitemOpen
  \bibfield  {author} {\bibinfo {author} {\bibfnamefont {Jeongwan}\ \bibnamefont {Haah}}, \bibinfo {author} {\bibfnamefont {Yunchao}\ \bibnamefont {Liu}}, \ and\ \bibinfo {author} {\bibfnamefont {Xinyu}\ \bibnamefont {Tan}},\ }\bibfield  {title} {\enquote {\bibinfo {title} {{ Efficient Approximate Unitary Designs from Random Pauli Rotations }},}\ }in\ \href {\doibase 10.1109/FOCS61266.2024.00036} {\emph {\bibinfo {booktitle} {2024 IEEE 65th Annual Symposium on Foundations of Computer Science (FOCS)}}}\ (\bibinfo  {publisher} {IEEE Computer Society},\ \bibinfo {address} {Los Alamitos, CA, USA},\ \bibinfo {year} {2024})\ pp.\ \bibinfo {pages} {463--475}\BibitemShut {NoStop}%
\bibitem [{\citenamefont {Metger}\ \emph {et~al.}(2024)\citenamefont {Metger}, \citenamefont {Poremba}, \citenamefont {Sinha},\ and\ \citenamefont {Yuen}}]{pseudorandom}%
  \BibitemOpen
  \bibfield  {author} {\bibinfo {author} {\bibfnamefont {Tony}\ \bibnamefont {Metger}}, \bibinfo {author} {\bibfnamefont {Alexander}\ \bibnamefont {Poremba}}, \bibinfo {author} {\bibfnamefont {Makrand}\ \bibnamefont {Sinha}}, \ and\ \bibinfo {author} {\bibfnamefont {Henry}\ \bibnamefont {Yuen}},\ }\bibfield  {title} {\enquote {\bibinfo {title} {{ Simple Constructions of Linear-Depth t-Designs and Pseudorandom Unitaries }},}\ }in\ \href {\doibase 10.1109/FOCS61266.2024.00038} {\emph {\bibinfo {booktitle} {2024 IEEE 65th Annual Symposium on Foundations of Computer Science (FOCS)}}}\ (\bibinfo  {publisher} {IEEE Computer Society},\ \bibinfo {address} {Los Alamitos, CA, USA},\ \bibinfo {year} {2024})\ pp.\ \bibinfo {pages} {485--492}\BibitemShut {NoStop}%
\bibitem [{\citenamefont {Mishmash}\ \emph {et~al.}(2023)\citenamefont {Mishmash}, \citenamefont {Gujarati}, \citenamefont {Motta}, \citenamefont {Zhai}, \citenamefont {Chan},\ and\ \citenamefont {Mezzacapo}}]{HCT_2023}%
  \BibitemOpen
  \bibfield  {author} {\bibinfo {author} {\bibfnamefont {Ryan~V.}\ \bibnamefont {Mishmash}}, \bibinfo {author} {\bibfnamefont {Tanvi~P.}\ \bibnamefont {Gujarati}}, \bibinfo {author} {\bibfnamefont {Mario}\ \bibnamefont {Motta}}, \bibinfo {author} {\bibfnamefont {Huanchen}\ \bibnamefont {Zhai}}, \bibinfo {author} {\bibfnamefont {Garnet Kin-Lic}\ \bibnamefont {Chan}}, \ and\ \bibinfo {author} {\bibfnamefont {Antonio}\ \bibnamefont {Mezzacapo}},\ }\bibfield  {title} {\enquote {\bibinfo {title} {Hierarchical clifford transformations to reduce entanglement in quantum chemistry wave functions},}\ }\href {\doibase 10.1021/acs.jctc.3c00228} {\bibfield  {journal} {\bibinfo  {journal} {Journal of Chemical Theory and Computation}\ }\textbf {\bibinfo {volume} {19}},\ \bibinfo {pages} {3194--3208} (\bibinfo {year} {2023})},\ \bibinfo {note} {pMID: 37227024}\BibitemShut {NoStop}%
\bibitem [{\citenamefont {Spencer}(1968)}]{SPENCER19681}%
  \BibitemOpen
  \bibfield  {author} {\bibinfo {author} {\bibfnamefont {Joel}\ \bibnamefont {Spencer}},\ }\bibfield  {title} {\enquote {\bibinfo {title} {Maximal consistent families of triples},}\ }\href {\doibase https://doi.org/10.1016/S0021-9800(68)80023-7} {\bibfield  {journal} {\bibinfo  {journal} {Journal of Combinatorial Theory}\ }\textbf {\bibinfo {volume} {5}},\ \bibinfo {pages} {1--8} (\bibinfo {year} {1968})}\BibitemShut {NoStop}%
\bibitem [{\citenamefont {Assmus}\ and\ \citenamefont {Key}(1992)}]{assmus_key_1992}%
  \BibitemOpen
  \bibfield  {author} {\bibinfo {author} {\bibfnamefont {E.~F.}\ \bibnamefont {Assmus}}\ and\ \bibinfo {author} {\bibfnamefont {J.~D.}\ \bibnamefont {Key}},\ }\href {\doibase 10.1017/CBO9781316529836} {\emph {\bibinfo {title} {Designs and their Codes}}},\ Cambridge Tracts in Mathematics\ (\bibinfo  {publisher} {Cambridge University Press},\ \bibinfo {year} {1992})\BibitemShut {NoStop}%
\bibitem [{\citenamefont {Feder}\ and\ \citenamefont {Subi}(2012)}]{feder2012packing}%
  \BibitemOpen
  \bibfield  {author} {\bibinfo {author} {\bibfnamefont {Tom{\'{a}}s}\ \bibnamefont {Feder}}\ and\ \bibinfo {author} {\bibfnamefont {Carlos~S.}\ \bibnamefont {Subi}},\ }\bibfield  {title} {\enquote {\bibinfo {title} {Packing edge-disjoint triangles in given graphs},}\ \ }(\bibinfo {year} {2012})\BibitemShut {NoStop}%
\bibitem [{\citenamefont {Bernien}\ \emph {et~al.}(2017)\citenamefont {Bernien}, \citenamefont {Schwartz}, \citenamefont {Keesling}, \citenamefont {Levine}, \citenamefont {Omran}, \citenamefont {Pichler}, \citenamefont {Choi}, \citenamefont {Zibrov}, \citenamefont {Endres}, \citenamefont {Greiner}, \citenamefont {Vuleti{\'{c}}},\ and\ \citenamefont {Lukin}}]{Bernien2017}%
  \BibitemOpen
  \bibfield  {author} {\bibinfo {author} {\bibfnamefont {Hannes}\ \bibnamefont {Bernien}}, \bibinfo {author} {\bibfnamefont {Sylvain}\ \bibnamefont {Schwartz}}, \bibinfo {author} {\bibfnamefont {Alexander}\ \bibnamefont {Keesling}}, \bibinfo {author} {\bibfnamefont {Harry}\ \bibnamefont {Levine}}, \bibinfo {author} {\bibfnamefont {Ahmed}\ \bibnamefont {Omran}}, \bibinfo {author} {\bibfnamefont {Hannes}\ \bibnamefont {Pichler}}, \bibinfo {author} {\bibfnamefont {Soonwon}\ \bibnamefont {Choi}}, \bibinfo {author} {\bibfnamefont {Alexander~S.}\ \bibnamefont {Zibrov}}, \bibinfo {author} {\bibfnamefont {Manuel}\ \bibnamefont {Endres}}, \bibinfo {author} {\bibfnamefont {Markus}\ \bibnamefont {Greiner}}, \bibinfo {author} {\bibfnamefont {Vladan}\ \bibnamefont {Vuleti{\'{c}}}}, \ and\ \bibinfo {author} {\bibfnamefont {Mikhail~D.}\ \bibnamefont {Lukin}},\ }\bibfield  {title} {\enquote {\bibinfo {title} {Probing many-body dynamics on a 51-atom quantum simulator},}\ }\href {\doibase 10.1038/nature24622} {\bibfield
  {journal} {\bibinfo  {journal} {Nature}\ }\textbf {\bibinfo {volume} {551}},\ \bibinfo {pages} {579--584} (\bibinfo {year} {2017})}\BibitemShut {NoStop}%
\bibitem [{\citenamefont {Bluvstein}\ \emph {et~al.}(2022)\citenamefont {Bluvstein}, \citenamefont {Levine}, \citenamefont {Semeghini}, \citenamefont {Wang}, \citenamefont {Ebadi}, \citenamefont {Kalinowski}, \citenamefont {Keesling}, \citenamefont {Maskara}, \citenamefont {Pichler}, \citenamefont {Greiner}, \citenamefont {Vuleti{\'{c}}},\ and\ \citenamefont {Lukin}}]{Bluvstein2022}%
  \BibitemOpen
  \bibfield  {author} {\bibinfo {author} {\bibfnamefont {Dolev}\ \bibnamefont {Bluvstein}}, \bibinfo {author} {\bibfnamefont {Harry}\ \bibnamefont {Levine}}, \bibinfo {author} {\bibfnamefont {Giulia}\ \bibnamefont {Semeghini}}, \bibinfo {author} {\bibfnamefont {Tout~T.}\ \bibnamefont {Wang}}, \bibinfo {author} {\bibfnamefont {Sepehr}\ \bibnamefont {Ebadi}}, \bibinfo {author} {\bibfnamefont {Marcin}\ \bibnamefont {Kalinowski}}, \bibinfo {author} {\bibfnamefont {Alexander}\ \bibnamefont {Keesling}}, \bibinfo {author} {\bibfnamefont {Nishad}\ \bibnamefont {Maskara}}, \bibinfo {author} {\bibfnamefont {Hannes}\ \bibnamefont {Pichler}}, \bibinfo {author} {\bibfnamefont {Markus}\ \bibnamefont {Greiner}}, \bibinfo {author} {\bibfnamefont {Vladan}\ \bibnamefont {Vuleti{\'{c}}}}, \ and\ \bibinfo {author} {\bibfnamefont {Mikhail~D.}\ \bibnamefont {Lukin}},\ }\bibfield  {title} {\enquote {\bibinfo {title} {A quantum processor based on coherent transport of entangled atom arrays},}\ }\href {\doibase
  10.1038/s41586-022-04592-6} {\bibfield  {journal} {\bibinfo  {journal} {Nature}\ }\textbf {\bibinfo {volume} {604}},\ \bibinfo {pages} {451--456} (\bibinfo {year} {2022})}\BibitemShut {NoStop}%
\bibitem [{\citenamefont {Baranyai}(1974)}]{baranyai1974factrization}%
  \BibitemOpen
  \bibfield  {author} {\bibinfo {author} {\bibfnamefont {Zsolt}\ \bibnamefont {Baranyai}},\ }\bibfield  {title} {\enquote {\bibinfo {title} {On the factrization of the complete uniform hypergraphs},}\ }\href@noop {} {\bibfield  {journal} {\bibinfo  {journal} {Infinite and finite sets}\ } (\bibinfo {year} {1974})}\BibitemShut {NoStop}%
\bibitem [{bei(2015)}]{beineke_wilson_2015}%
  \BibitemOpen
  \href {\doibase 10.1017/CBO9781139519793} {\emph {\bibinfo {title} {Topics in Chromatic Graph Theory}}},\ Encyclopedia of Mathematics and its Applications\ (\bibinfo  {publisher} {Cambridge University Press},\ \bibinfo {year} {2015})\BibitemShut {NoStop}%
\bibitem [{\citenamefont {Alspach}\ \emph {et~al.}(1990)\citenamefont {Alspach}, \citenamefont {Bermond},\ and\ \citenamefont {Sotteau}}]{hpath_cycle}%
  \BibitemOpen
  \bibfield  {author} {\bibinfo {author} {\bibfnamefont {B.}~\bibnamefont {Alspach}}, \bibinfo {author} {\bibfnamefont {J.-C.}\ \bibnamefont {Bermond}}, \ and\ \bibinfo {author} {\bibfnamefont {D.}~\bibnamefont {Sotteau}},\ }\enquote {\bibinfo {title} {Decomposition into cycles i: Hamilton decompositions},}\ in\ \href {\doibase 10.1007/978-94-009-0517-7_2} {\emph {\bibinfo {booktitle} {Cycles and Rays}}}\ (\bibinfo  {publisher} {Springer Netherlands},\ \bibinfo {address} {Dordrecht},\ \bibinfo {year} {1990})\ pp.\ \bibinfo {pages} {9--18}\BibitemShut {NoStop}%
\bibitem [{\citenamefont {Glebov}\ \emph {et~al.}(2017)\citenamefont {Glebov}, \citenamefont {Luria},\ and\ \citenamefont {Sudakov}}]{hpath_graph}%
  \BibitemOpen
  \bibfield  {author} {\bibinfo {author} {\bibfnamefont {Roman}\ \bibnamefont {Glebov}}, \bibinfo {author} {\bibfnamefont {Zur}\ \bibnamefont {Luria}}, \ and\ \bibinfo {author} {\bibfnamefont {Benny}\ \bibnamefont {Sudakov}},\ }\bibfield  {title} {\enquote {\bibinfo {title} {The number of hamiltonian decompositions of regular graphs},}\ }\href {\doibase 10.1007/s11856-017-1583-y} {\bibfield  {journal} {\bibinfo  {journal} {Israel Journal of Mathematics}\ }\textbf {\bibinfo {volume} {222}},\ \bibinfo {pages} {91--108} (\bibinfo {year} {2017})}\BibitemShut {NoStop}%
\bibitem [{\citenamefont {Galicia}\ \emph {et~al.}(2020)\citenamefont {Galicia}, \citenamefont {Ramon}, \citenamefont {Solano},\ and\ \citenamefont {Sanz}}]{hpath_quantum}%
  \BibitemOpen
  \bibfield  {author} {\bibinfo {author} {\bibfnamefont {Asier}\ \bibnamefont {Galicia}}, \bibinfo {author} {\bibfnamefont {Borja}\ \bibnamefont {Ramon}}, \bibinfo {author} {\bibfnamefont {Enrique}\ \bibnamefont {Solano}}, \ and\ \bibinfo {author} {\bibfnamefont {Mikel}\ \bibnamefont {Sanz}},\ }\bibfield  {title} {\enquote {\bibinfo {title} {Enhanced connectivity of quantum hardware with digital-analog control},}\ }\href {\doibase 10.1103/PhysRevResearch.2.033103} {\bibfield  {journal} {\bibinfo  {journal} {Phys. Rev. Res.}\ }\textbf {\bibinfo {volume} {2}},\ \bibinfo {pages} {033103} (\bibinfo {year} {2020})}\BibitemShut {NoStop}%
\bibitem [{\citenamefont {Sidon}(1932)}]{Sidon1932}%
  \BibitemOpen
  \bibfield  {author} {\bibinfo {author} {\bibfnamefont {S.}~\bibnamefont {Sidon}},\ }\bibfield  {title} {\enquote {\bibinfo {title} {Ein satz {\"u}ber trigonometrische polynome und seine anwendung in der theorie der fourier-reihen},}\ }\href {\doibase 10.1007/BF01455900} {\bibfield  {journal} {\bibinfo  {journal} {Mathematische Annalen}\ }\textbf {\bibinfo {volume} {106}},\ \bibinfo {pages} {536--539} (\bibinfo {year} {1932})}\BibitemShut {NoStop}%
\bibitem [{\citenamefont {Golomb}(1972)}]{golomb1972number}%
  \BibitemOpen
  \bibfield  {author} {\bibinfo {author} {\bibfnamefont {Solomon~W}\ \bibnamefont {Golomb}},\ }\bibfield  {title} {\enquote {\bibinfo {title} {How to number a graph, graph theory and computing},}\ }\href@noop {} {\bibfield  {journal} {\bibinfo  {journal} {Academic Press, New York}\ }\textbf {\bibinfo {volume} {23}},\ \bibinfo {pages} {37} (\bibinfo {year} {1972})}\BibitemShut {NoStop}%
\bibitem [{\citenamefont {Drakakis}(2009)}]{drakakis2009review}%
  \BibitemOpen
  \bibfield  {author} {\bibinfo {author} {\bibfnamefont {Konstantinos}\ \bibnamefont {Drakakis}},\ }\bibfield  {title} {\enquote {\bibinfo {title} {A review of the available construction methods for golomb rulers},}\ }\href {\doibase 10.3934/amc.2009.3.235} {\bibfield  {journal} {\bibinfo  {journal} {Advances in Mathematics of Communications}\ }\textbf {\bibinfo {volume} {3}},\ \bibinfo {pages} {235--250} (\bibinfo {year} {2009})}\BibitemShut {NoStop}%
\end{thebibliography}%
\end{document}